\documentclass[acmsmall, authorversion,nonacm]{acmart}

\AtBeginDocument{%
  \providecommand\BibTeX{{%
    \normalfont B\kern-0.5em{\scshape i\kern-0.25em b}\kern-0.8em\TeX}}}

\usepackage{booktabs}
\usepackage{caption}
\usepackage{subcaption}
\usepackage{color, colortbl}
\definecolor{LightCyan}{rgb}{0.88,1,1}
\begin{document}

\title[]{Designing a User-centric Framework for Information Quality Ranking of Large-scale Street View Images}
\author{Tahiya Chowdhury}
\authornote{Both authors contributed equally to this research.}
\authornote{Work done while at Rutgers University.}
\affiliation{%
  \institution{Colby College}
  \city{Maine}
  \country{USA}}
\email{tahiya.chowdhury@colby.edu}

\author{Ilan Mandel}
\authornotemark[1]
\affiliation{%
  \institution{Cornell Tech}
\city{New York}
  \country{USA}}
\email{im334@cornell.edu}

\author{Jorge Ortiz}
\affiliation{%
  \institution{Rutgers University}
  \city{New Jersey}
  \country{USA}}
\email{jorge.ortiz@rutgers.edu}

\author{Wendy Ju}
\affiliation{%
  \institution{Cornell Tech}
  \city{New York}
  \country{USA}}
\email{wendyju@cornell.edu}

\renewcommand{\shortauthors}{Chowdhury and Mandel, et al.}

\begin{abstract}
Street view imagery (SVI), largely captured via outfitted fleets or mounted dashcams in consumer vehicles is a rapidly growing source of geospatial data used in urban sensing and development. These datasets are often collected opportunistically, are massive in size, and vary in quality which limits the scope and extent of their use in urban planning.
Thus far there has not been much work to identify the obstacles experienced and tools needed by the users of such datasets. This severely limits the opportunities of using emerging street view images in supporting novel research questions that can improve the quality of urban life. This work includes a formative interview study with 5 expert users of large-scale street view datasets from academia, urban planning, and related professions which identifies novel use cases, challenges, and opportunities to increase the utility of these datasets.
Based on the user findings, we present a framework to evaluate the quality of information for street images across three attributes (spatial, temporal, and content) that stakeholders can utilize for estimating the value of a dataset, and to improve it over time for their respective use case.
We then present a case study using novel street view images where we evaluate our framework and present practical use cases for users. We discuss the implications for designing future systems to support the collection and use of street view data to assist in sensing and planning the urban environment.


\end{abstract}


\maketitle


\section{Introduction}

Street view imagery (SVI) has become a practical source of large-scale image data to sense the built environment and serves a crucial role in understanding urban landscapes. SVI features large numbers of images taken along streets from different locations, providing scenes captured from the perspective of a human. Initially conceptualized for mapping the streets to allow people to visit a place without physical travel, SVI has been used in many research studies and applications for both qualitative and quantitative assessment of the physical environment. Advances in image processing techniques for tasks such as object detection, segmentation, and scene classification, along with dramatically cheaper digital storage, have increased the accessibility and utility of these large street view datasets. 

Unlike aerial imagery, which can be acquired via satellites, aircraft and drones, SVI is captured with on-ground mobile devices, most typically, via automobiles. SVI data is suitable for extracting visual features that can be of interest for urban computing studies estimating characteristics of the built environment such as building age and style~\cite{li2018estimating, hoffmann2019model, GONZALEZ2020106805}, neighborhood quality of life~\cite{bin2020multi, doi:10.1021/acs.est.0c05572,
weld2019deep, zhang2018walking} and pedestrian interaction in space~\cite{chen2020estimating, wang2020relationship, goel2018estimating} among the many other use cases. SVI is currently available through both commercial services that rely on the systematic protocol for frequent data collection and crowd-sourcing services that rely on volunteer-contributed images at irregular frequency. While a systematic data acquisition schedule promises temporal consistency, SVI services like Google Street View (GSV) lack temporal freshness (e.g. not updated frequently) and spatial coverage (e.g. GSV cars cannot cover gated neighborhoods unless explicit permission is obtained) offered through services relying on crowd-sourcing. Real-time data acquisition and dense mapping thus bring more value for researchers interested in studying large-scale, multiple geographic regions during different seasons or a certain time of the day. However, similar to other types of real-world data collection via crowd contribution, SVI suffers from information quality issues due to device \textit{heterogeneity}, \textit{incompleteness}, and \textit{inconsistency}.

\begin{figure}
\centering
    \includegraphics[width=.49\linewidth]{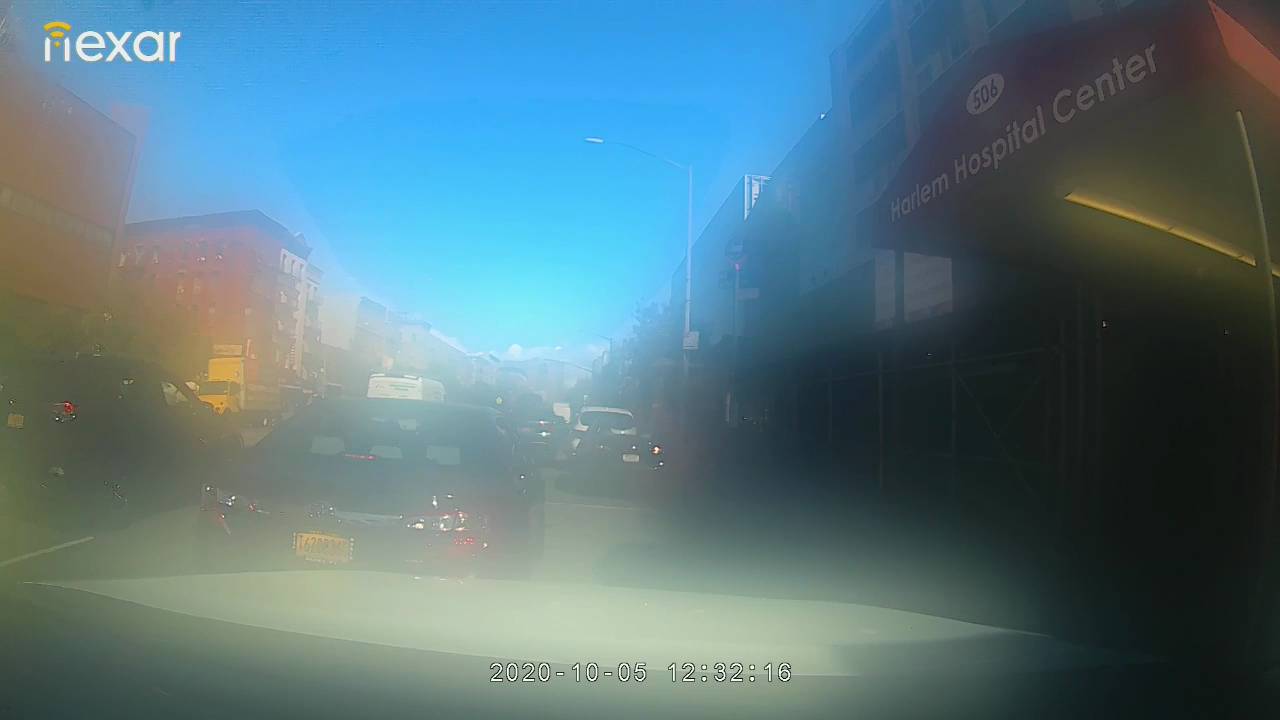}
    \includegraphics[width=.495\linewidth]{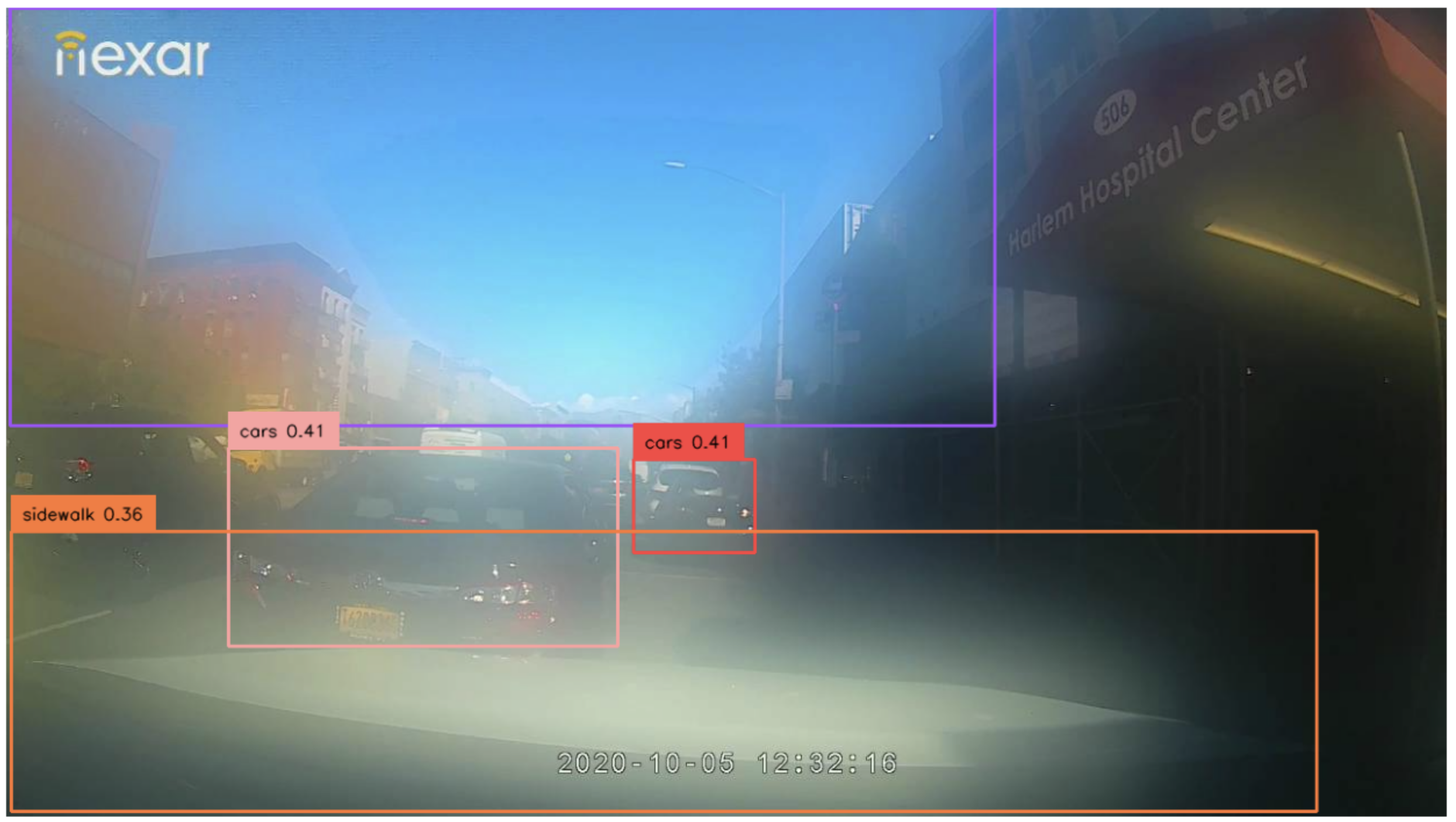}\hfill
    \caption{A blurry image example from Nexar dashboard camera image dataset (left). The blurriness would lead to low quality of information for many applications that seeks to extract insights from the content of the image (i.e. number of cars or sidewalk width), but can be of sufficient quality for sky view estimation applications.}
\label{fig:nexar_low_QoI}
\end{figure}

Consider an example street view image from a dashboard camera shown in Figure~\ref{fig:nexar_low_QoI}. While a human observer can observe the general attributes from the image (e.g. proximity to a hospital building, sky view), even advanced image processing techniques will fail to estimate distance (e.g. sidewalk width) or building heights due to the blurriness of the image. As more services are emerging providing street view footage with dashboard cameras and other devices, novel large-scale SVI data can open up many analysis and application paradigms. However, high-quality street-level mapping requires petabytes (1 petabyte $\approx 10^6$ gigabytes) to store the images and large-scale computing infrastructure to analyze and extract information. The size and quality variance of these datasets create a fundamental limit on their utility to all but expert users. 
Improving assesments of data quality can lead to more efficient analysis and lower the environmental and financial costs associated with storage and computing needs. While researchers and data providers are introducing new functionalities and applications using SVI services, there has been little research on identifying prospective users of these datasets for urban planning. A key research gap remains in identifying the challenges and opportunities of SVI from the perspective of urban planners who use them as a proxy to estimate the quality of urban life such as through virtual site audits.

In this work, our focus is specifically on 
how novel SVI can open up new research opportunities and applications in urban planning and recognizing obstacles faced by professionals in studying different aspects of urban living. We focus on challenges and use cases related to information quality in assessing large-scale street view dataset's value, particularly for urban planning and data analysis applications. For a comprehensive understanding of the experience of users, we draw on interviews with 5 academics and professionals who regularly have used one or more SVI datasets in their data analysis and planning workflow. Findings from these interviews highlight the ways in which urban planners and researchers currently utilize street view images in their work, what factors they use to make such decisions, the informational challenges they face, and the potential solutions to mitigate that. Our analysis of the interviews reveals that lack of interactive tools, quality assessment of data based on its analysis value, and cost model are the primary barriers that limit the ways an urban planner navigates through their interaction with the datasets.

To address these challenges, we identified three quality attributes based on our interview study for measuring SVI information quality: spatial, temporal, and content quality, and designed a framework to describe the quality of an SVI dataset in relation to other data segments. We evaluated these attributes and the unified quality assessment framework using a novel SVI dataset collected in New York City by vehicle instrumented with dashboard cameras. This article makes three key contributions to the field of sustainable computing. First, we contribute nuanced empirical understandings of the innovative use cases and challenges for street view images, which provide the first of its kind stakeholder's perspectives on the use of these datasets. Second, we contribute a quality evaluation framework for assessing the quality of information and ranking, which extends existing work on SVI quality~\cite{HOU2022103094} by developing a single quality metric for improving data over time. Third, we synthesize our findings to highlight design trade-offs and considerations for enhancing future SVI-based urban computing systems to support user-centric human-AI collaboration through sustainable data practices.

\section{Background and Related Work}\label{sec:background}
Our work is informed by existing SVI services and their features, use, and challenges of SVI in urban computing as well as prior work on data quality for such imagery.

\subsection{SVI Services and Their Features}

Google Street View (GSV) is the most frequently used street view image dataset. Originally based on the Stanford City Block project, it first launched in several cities in 2007 \cite{olanoff2013inside}. Currently, GSV imagery is estimated to cover half of the human population~\cite{goel2018estimating} and is active in more than 100 countries~\cite{BILJECKI2021104217}. GSV images are collected with car-mounted 360-degree panoramic cameras for roads and have recently expanded backpack-mounted cameras to cover landmarks and indoor spaces~\cite{5481932}. They have a well-developed interface with a pay-as-you-go access model, omni-directional footage, and standardized collection procedure. Images are collected at a semi-regular frequency, generally in the daytime under sufficient lighting and good weather conditions to ensure quality. Google publishes online which regions they are currently collecting data from. Several instances of GSV-like services have emerged such as Tencent Street View and Baidu Total View (China), Bing Maps Streetside, Cyclomedia (Europe), Yandex (Russia), and Apple's Look Around which also provide access to street view imagery.

\begin{figure}[t]
     \centering
          \begin{subfigure}[b]{0.3\textwidth}
         \centering
         \includegraphics[width=\textwidth]{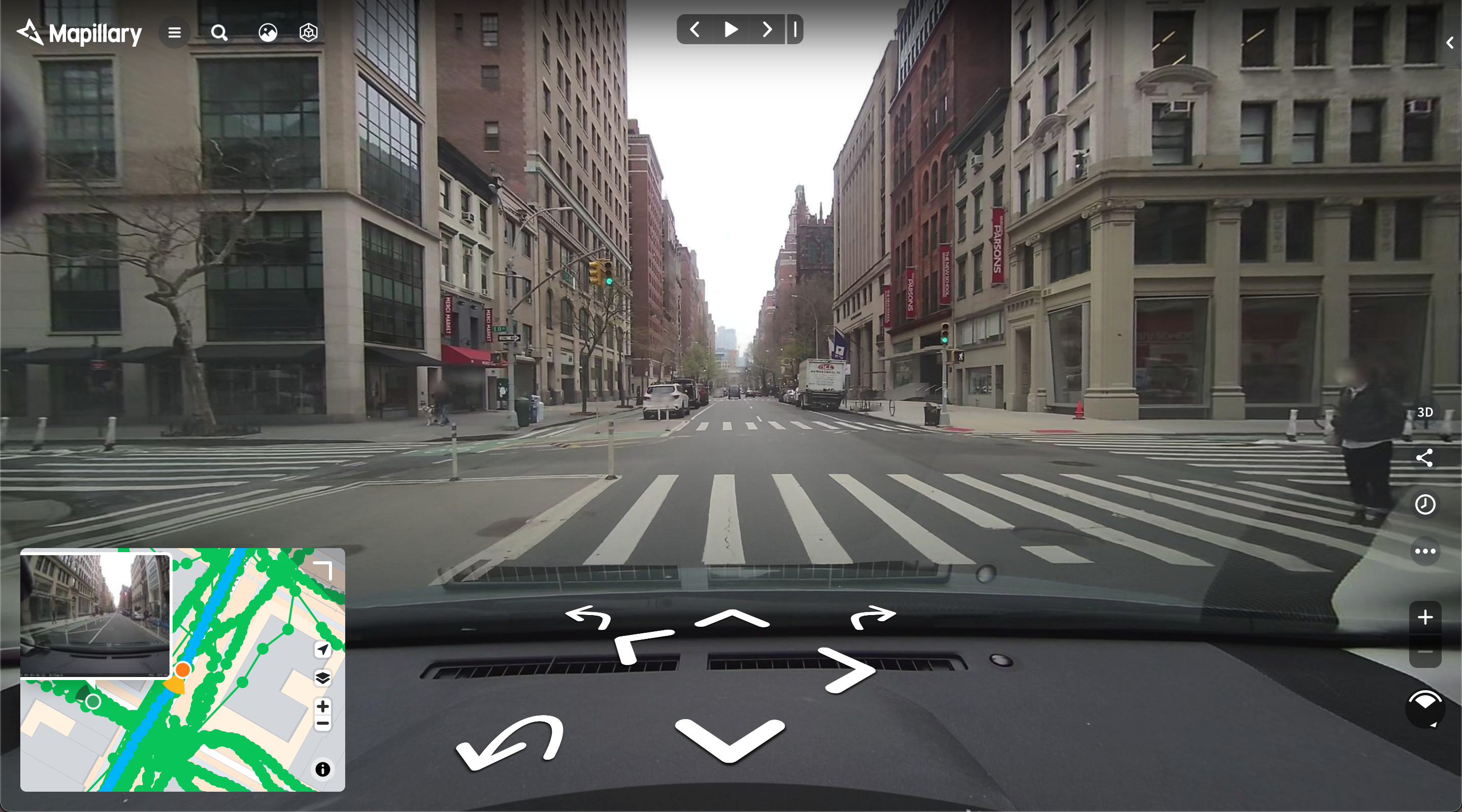}
         \caption{Mapillary}
         \label{fig:mapillary_ex}
     \end{subfigure}
     \hfill
     \begin{subfigure}[b]{0.3\textwidth}
         \centering
         \includegraphics[width=\textwidth]{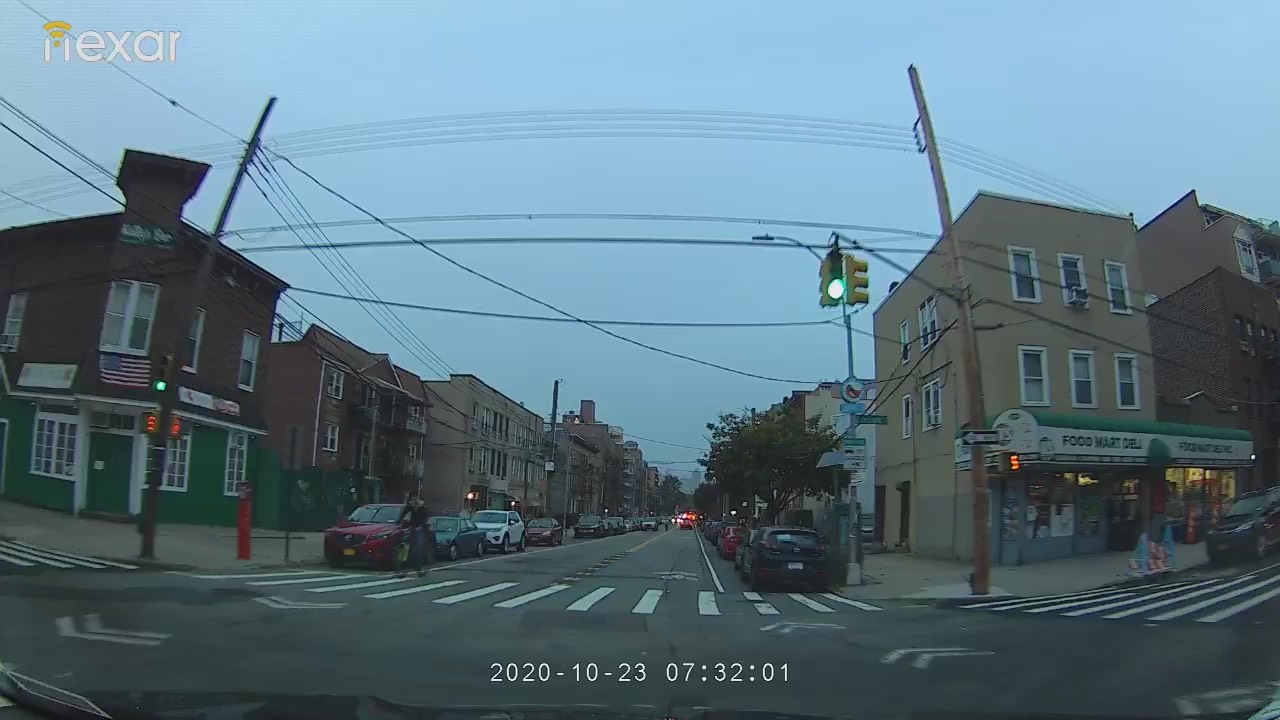}
         \caption{Nexar}
         \label{fig:nexar_ex}
     \end{subfigure}
     \hfill
     \begin{subfigure}[b]{0.3\textwidth}
         \centering
         \includegraphics[width=\textwidth, trim={0 0 0 6.5cm},clip]{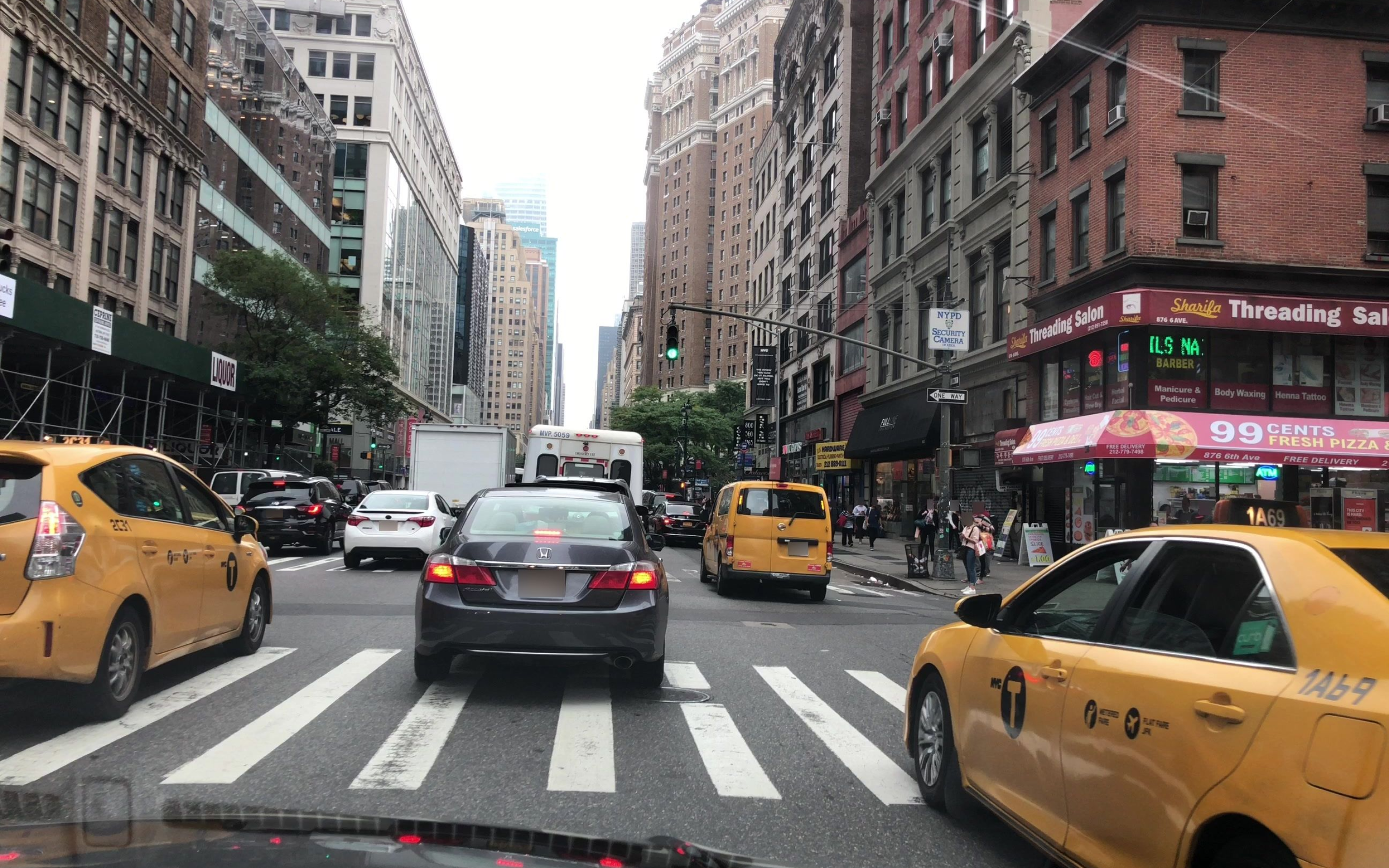}
         \caption{KartaView}
         \label{fig:karta_ex}
     \end{subfigure}
     \caption{Example images from various street view datasets other that Google Street View}
     \label{streetview_examples}
\end{figure}

There are several services that provide commercial access to street-view images collected through crowd-sourcing. Two such services, Mapillary and KartaView, have become popular as users can capture video or images with any device and contribute. Unlike GSV where locations are typically revisited approximately once a year, crowd-sourced images can be of much higher temporal resolution, especially for areas frequently visited by many people. However, as image collection does not follow a systematic protocol, images are captured using various devices and camera orientations, in varying weather, lighting conditions, and different times of the day and year~\cite{neuhold2017mapillary}. As a result, data suffer from heterogeneous quality issues. Spatial and temporal coverage of an area is variable, as the locations captured tend to be opportunistic rather than strategic. As a result, the mapping of a geographic area can be less comprehensive than that of GSV but is updated more often. 

Recently, Automotive Original Equipment Manufacturers (OEMs) such as Nexar have emerged as a street view imagery service. Nexar collects its street view imagery via network-connected dashboard cameras for insurance claims made by drivers. Images are automatically reported at regular intervals to a central cloud server. Street view images collected in this automated protocol offer higher temporal resolution similar to crowd-sourced services in comparison but with less heterogeneity. Nexar thus generates real-time road data that provides a digital twin of the geographic regions by processing images using artificial intelligence-enabled object detection and other automated vision techniques to outline work zones, road assets, and parking. Figure~\ref{streetview_examples} shows example images from a number of street view image data sources.

\subsection{Street View Image Datasets Uses and Applications}

Street view images open opportunities for urban researchers and policymakers to use the images as a virtual representation of street scenes in the real world. Local governments might need regular data on a large geographic area while community groups may need higher fidelity data on only a few blocks. On the othe rhand, a real estate developer may need specific measurements of a singular block.  Street View Imagery is thus increasingly being used~\cite{biljecki2021street} along with sensor networks~\cite{rashid2016applications}, satellite imagery~\cite{miller2003cities}, urban ethnography \cite{jackson1985urban} and survey/census data~\cite{logan2018relying} to understand the facets of urban life.   
Urban planners have used street view images as a proxy for traditional survey methods so that researchers can analyze street activities without directly observing the streets to collect data \cite{shapiro2018street, vandeviver2014applying}. 

\subsubsection{Street View Images in Urban Computing} \label{sec:background_urban_computing}

Collecting data at a city scale is a time-consuming and expensive, yet an essential step for urban analytics. GSV images have been used by researchers to classify urban street features such as trees~\cite{thirlwell2020big, 8930566}, utility poles~\cite{krylov2018automatic}, traffic signs~\cite{campbell2019detecting, 10.1145/3180496.3180638}, bike racks~\cite{maddalena2020mapping}, potholes and other defects~\cite{9072401},  manholes~\cite{vishnani2020manhole} and sidewalks~\cite{doi:10.1177/2399808321995817} in an effort to map their distribution across the city. Similarly, the images have been used to estimate road quality and classify road surface types~\cite{zhang2020automatic, doi:10.1177/2399808319828734}, land use and urban zones (residential, industrial, etc.)~\cite{feng2018urban, chang2020mapping}, and commercial activity~\cite{10.1145/3184558.3186581, 8890683}. Building level static attributes such as building age~\cite{li2018estimating}, heights, use, and architectural style~\cite{hoffmann2019model, GONZALEZ2020106805}, and energy consumption~\cite{guhathakurta2020spatial} have been estimated from street view images.

\begin{figure*}
 \includegraphics[width=0.46\linewidth]{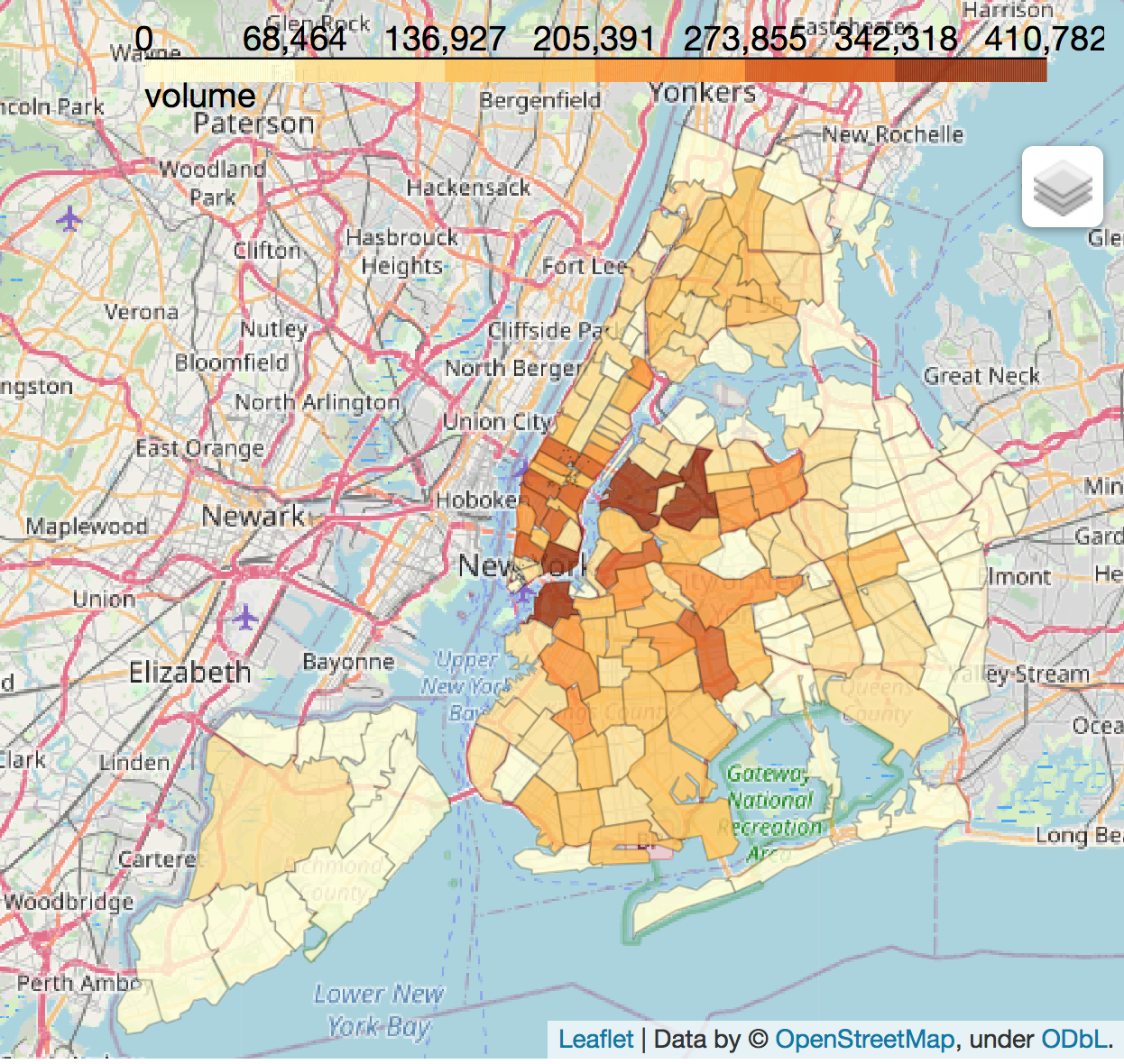}
  \includegraphics[width=0.48\linewidth]{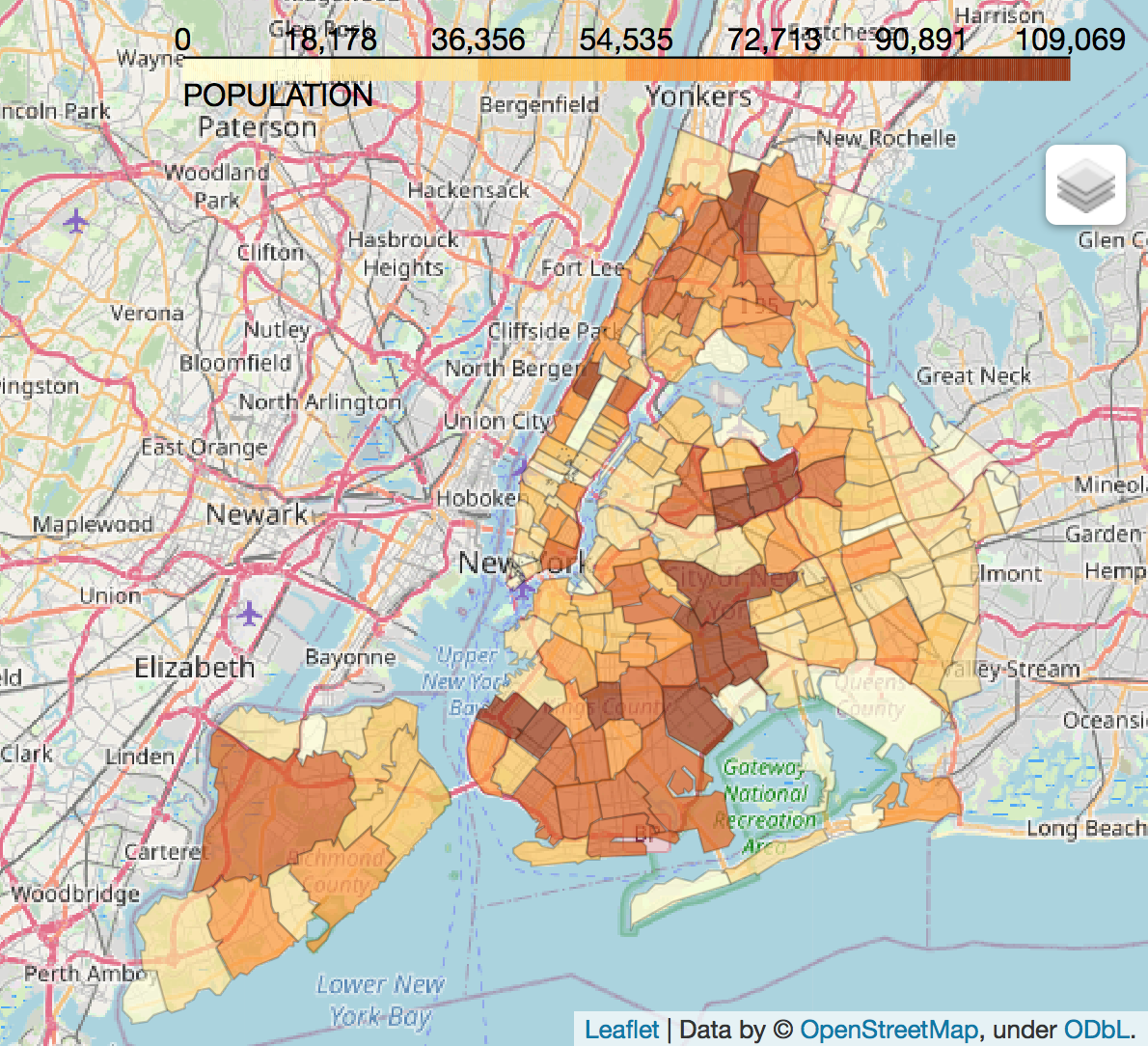}
  \caption{Data distribution of an image dataset collected by vehicles instrumented with dashboard cameras across 196 Zip code areas in New York City (left). Population of each Zip area is shown on right for reference. Note that representation in the dataset does not commensurate the population in most cases.}
  \label{fig:data volume vs population of the city}
  \vspace{-3mm}
\end{figure*}

Attributes such as number of lanes~\cite{cicchino2020not}, sidewalk width and surface~\cite{isola2019google}, car driving difficulty \cite{skurowski2018saliency},  pedestrian volume \cite{chen2020estimating}, mobility patterns and route choices~\cite{wang2020relationship, goel2018estimating, verhoeven2018differences} have been estimated from street view images to explore their relationship to pedestrian and drivers safety. Researchers studying urban health have studied the relationship between the built environment and urban occupants' exposure to dengue fever \cite{andersson2019combining} and COVID-19 \cite{ijerph17176359} flu. In recent years, SVIimages have been used to quantify sun glare, sky view, light pollution, shade percentage~\cite{gong2018mapping, liang2020gsv2svf, zeng2018fast}, greenery, walkability, and physical attractiveness to study their links to various health outcomes such as happiness~\cite{hart2018contextual}, obesity and physical activity ~\cite{hart2018contextual}, air quality~\cite{doi:10.1021/acs.est.0c05572}, and mental health~\cite{wang2019urban}. Massive spatial coverage of street view images has been useful in studying urban walkability and accessibility based on crowdedness~\cite{cao2018walkway, weld2019deep, zhang2018walking}, property value based on neighbourhood amenities~\cite{law2019take, bin2020multi, fu2019street}, and predict socio-economic factors such as income~\cite{diou2018image} and crime rates~\cite{Vandeviver2014ApplyingGM}.

Computer vision based recognition along with high performance computing allows processing of large collection of images at scale to sense changes over a long period to compare similar attributes in different cities, that makes "social sensing" possible, through analysis of how people's use of the city ebbs and flows over time~\cite{liu2015social}. Fixed public web cameras, often limited to landmarks and points of interests, have been used to estimate pedestrian movement patterns \cite{slobogin2002public, hipp2016webcams, seer2014kinects}, pedestrian volume \cite{liu2009pedestrian, petrasova2019visualization} and social distancing behavior~\cite{casalmoore_2020, chowdhury2021towards} over time. Vehicle-mounted cameras, on the other hand, can provide large area coverage and temporal frequency that have been useful to estimate the spatio-temporal distribution of taxi trips~\cite{zhang2019social} and predict pedestrian intent \cite{kataoka2015fine}.  


\subsubsection{Tools for interacting with street view images} \label{sec:background_human_centered}



Many urban planning studies have traditionally relied on field visits and audits to tackle questions concerning the built environment's impact on public health, perception of safety, etc.~\cite{RePEc:sae:urbstu:v:45:y:2008:i:9:p:1973-1996, sallis2012role, CLIFTON200795}. Designing creative, interactive tools to support the works of professionals for the exploration of urban built environments has garnered attention in the HCI community. Earlier works~\cite{10.1145/1518701.1518871, 10.1145/302979.303114} introduced systems to integrate field visit data with existing maps on a unified platform for ease of comprehension and analysis. The emergence of street view images and progress in computer vision has enabled processing and querying large-scale image data sets for virtual audits~\cite{10.1145/2647868.2654948, https://doi.org/10.48550/arxiv.1404.1777, lowe2004distinctive}, scaling up urban planning with wider coverage, as a substitute for time-consuming and labor-intensive physical site visits. Following the practice of incorporating stakeholders' input in urban design processes~\cite{10.1145/3173574.3173769, 10.1145/3290605.3300571}, \cite{10.1145/3290605.3300292} used crowd-sourcing to label accessibility issues on GSV images. However, large-scale virtual auditing requires expert input in labeling that street users often lack and auditors need tools and methods to visually explore large-scale image data. A number of visual analytical tools have been introduced to explore spatio-temporal data based on measures of air pollution~\cite{4376168}, mobility~\cite{6876029, 6634127, 7539380}, accessibility~\cite{miranda2020urban} and transportation patterns~\cite{10.1145/2557500.2557537}. 

\subsubsection{Data Governance for Street View Images}

SVI presents legitimate privacy concerns as they are captured in public spaces and without consent, can be made available to anyone with internet access, and can be used for malicious attempts such as stalking, facilitating crimes, and compromising privacy~\cite{Rakower}. 
It has become the norm to automatically blur faces and license plates~\cite{35481} or remove pedestrians and vehicles using computer vision \cite{https://doi.org/10.48550/arxiv.1903.11532, 5543255, nodari2012digita} to maintain privacy. Individuals' street-level behavior can reveal sensitive details, such as who someone is, what they are doing, and who accompanied them at a particular place and time \cite{rakower2011blurred}.

\subsection{Quality of Information for Street View Images}

Images generate 'general-purpose' information that may be applied to many use cases. Images collected in a non-standardized process can be deemed of sufficient quality for some applications while needing a higher quality for others due to heterogeneous quality in device orientation, type, and view completeness. Heuristics such as collecting more information or maximizing coverage does not guarantee improved data quality and better decision-making (e.g. air quality or temperature changes slowly, so higher sampling frequency may not add more value to the data in regular environmental sensing applications).  

Quality of Information in a sensor network is a metric used to evaluate information quality to make decisions about the utility of information~\cite{10.1145/2489253.2489265}. It is defined as the difference between observed measurements of an event via sensing and the actual measurement~\cite{4660116}. It is closely related to Value of Information~\cite{article12}, the utility of information for a specific application, and Age of Information~\cite{yates2021age}, the timeliness of information delivery. 
The quality of information can be regarded as context-independent whereas the value of information is context-dependent. Even high information quality can offer low information value for a specific application~\cite{4144841} (e.g. temperature data at minute resolution for transit design). In this work, we focus on information quality to provide a general framework for quality evaluation that can be tailored for specific applications and users.



Prior work has designed frameworks for scoring and ranking an information segment based on spatio-temporal context, accessibility, trustworthiness, data provenance, timeliness, presentation, and relevance as key characteristics to evaluate application-specific utility. ~\citet{10.1145/2489253.2489265} introduced a two-layer QoI/VoI framework
for scoring and ranking information products based on their QoI and VoI attributes taxonomy. Along a similar goal,~\citet{6935003} proposed sensor selection based on application-specific QoI, and~\citet{7295461} used QoI for dynamic data delivery based on user query. As mobile sensors cover more area over a period of time than an equal number of stationary sensors~\cite{10.1145/1062689.1062728}, the quality of data coverage achieved by mobile sensors depends on the velocity, mobility pattern, and the number of mobile sensors deployed as explored in~\cite{10.1145/1161089.1161102}.~\cite{BALLARI2012102, NGAI2014203} explored adaptive QoI-aware data sampling in mobile sensor networks for optimizing the value and cost of information. To address varying spatial and temporal coverage, \citet{wang2006movement} proposed algorithms to detect missing holes in coverage to identify sampling locations and improve coverage.

\subsubsection{Quality of Information for Street View Imagery.} 
Most works evaluating SVI are focused primarily on spatial coverage. \citet{juhasz2016user} compared spatial coverage of GSV and Mapillary, and reported Mapillary to provide better coverage over time for less visited road segments. ~\citet{ma2019state} reported crowdsourced SVI to be skewed to specific geographic regions and seasons due to reliance on user contribution, which was also observed by ~\citet{mahabir2020crowdsourcing}. ~\citet{fry2020assessing} specifically studied image contribution patterns and found that regions with higher socio-economic conditions are sampled more frequently, which can induce bias in the downstream applications. 

~\citet{HOU2022103094} identified notable quality issues reported in prior SVI literature and developed an extensive quality framework including image quality (blurriness, lighting), redundancy, spatial and temporal coverage, privacy, and metadata availability (location, timestamp). Our contribution is most closely related to this work as we also focus on developing a scoring metric to evaluate the information quality of SVI. The key difference is we measure the information quality of a data segment with the goal to rank different data segments, whereas~\citet{HOU2022103094} focused on visualization. While visualization is useful when exploring a specific region (e.g. zip code area), a ranking can help users to compare and select certain areas from many available ones or data from a particular service from the many options covering that area. As newer services for SVI data start becoming available (e.g. on-vehicle footage from Nexar, Mobileye), it is important to get in touch with stakeholders who use these data to understand their workflow and missing links that existing services, tools, and applications fail to deliver. Finally, many works have looked at developing tools and applications based on AI-based information extraction. Our framework also considers performance on common computer vision tasks (object detection, semantic segmentation) for image analysis and incorporates that into the proposed information quality framework.

\section{Formative Study: Expert Interviews}

We conducted expert interviews with researchers and designers in urban planning, and analytics to understand the uses and challenges for street view images in their current work practices. Our study was approved by Cornell Institutional Review Board.
We first describe the details of the interviews (participants, procedure, data analysis) and then present the findings from these interviews regarding how street view image datasets are utilized in the participants' workflow, what challenges they encounter in using such datasets, and what novel SVI features can enable more insights.


\subsection{Participants}

We performed semi-structured interviews with 5 professionals and academics who use street view images in their professional work (3 female, 2 male). Participants were recruited through our research network and snowball sampling. During the interviews, we grounded our discussion with the participants using a street view image dataset as an exemplar for demonstration purposes that was collected in New York City. We specifically recruited participants who live and work in The Greater New York Area to ensure that our interviews reflect both their professional and personal perspectives to identify specific needs and contextual examples. Most participants have experience with one or more Street View image services for studying urban environments at various scales, from single sites to neighborhoods to entire cities. Participants were primarily familiar with Google Street View images and were intentionally selected to understand varied professional backgrounds. Many of them work closely with computing professionals to develop tools for their uses of SVI datasets. 

We performed interviews with: 
\begin{itemize}
    \item \textit{P1:} A senior executive and a research scientist from the NYC Department for the Aging (DFTA).
    \item \textit{P2:} An Urban Designer and Civic tech entrepreneur who is experienced in building community-centered tools for the NYC Department of City Planning.
    \item \textit{P3:} An urban planning executive for the Flatiron Nomad Partnership Business Improvement District (BID).
    \item \textit{P4:} An assistant professor of Civil and Mechanical Engineering focused on using mobile sensing technologies to monitor urban and structural environments.
    \item \textit{P5:} An assistant professor of urban informatics with a focus on urban energy, land use, and mobility.
\end{itemize}

\subsection{Procedure}

We conducted the interviews remotely through a video conferencing tool (Zoom). All interviews were conducted by two of the authors simultaneously between December 2021 and March 2022. At the beginning of each interview, we collected verbal consent from participants to record the interview. Interviews lasted for 48--52 minutes. Video recordings of the interviews were transcribed using audio transcription available on Zoom, which was further edited for capitalization and punctuation if not captured by the transcript.Interviews covered the expert's background, past usage of street view datasets, and views on potential usage of data within their own professional work. 

The interviews are conducted in a semi-structured format so that participants can engage in spontaneous conversations about their experiences of using and interacting with street view image services. Our broader interview protocol focused on understanding how urban planners and researchers use street view images as a source of visual information about the city, what are the potential uses, and the limitations of currently available datasets. To understand the expert's perspectives with respect to their work context, we also asked about the participant's background that may require gathering visual insights about the city and their past usage of SVI and other data sources for collecting these insights. To ground the discussion, we showed participants images from a large-scale data-set of dashcam data collected by Nexar, shown in Figure \ref{fig:nexar_ex}.




\subsection{Data Analysis}
We followed a thematic analysis procedure for data analysis~\cite{doi:10.1191/1478088706qp063oa, khandkar2009open, braun2006using}. We began by having the two authors read the interview transcripts in asynchronous sessions and open-code the interview transcript for emergent themes. The primary focus of the coding process was to identify the expert's perspective on the challenges encountered in using street view images for urban planning. Next, the two first authors synchronously met to re-read the codes, write analytic memos on the codes identified individually, and collate them into preliminary themes through an iterative process of comparing data to themes. Finally, we organized the themes according to the challenges urban planners experience in their interaction with street-view images. The themes capture the participant's workflow for gathering insights, what informational challenges hinder the process, what additional tools or information can help resolve these complexities, and the potential solution that could improve their experience.

\section{Formative Study: Findings}\label{sec:needfinding}

Street view images are complex heterogeneous datasets that contain useful information for many urban applications. They can become a proxy for virtual audits of urban physical surveys and can be used for developing insights about different static and dynamic information in the urban landscape. Unlike most sensing modalities, an image contains general-purpose information and can add ambiguity to its interpretation. High-resolution images require large-scale storage, which can often create hindrances for stakeholders who are interested in using the datasets. Many stakeholders look for tools requiring less technical expertise so that they can utilize the dataset through a blend of their disciplinary knowledge and interactive explorations. The heterogeneous quality can exacerbate the challenges associated with storage and exploration. Below we detail the key findings of our interview for potential future design choices.

\subsection{Uses and limitations of current Street View Image}

\subsubsection{Uses.} To understand a neighborhood's socio-economic, health, political, and accessibility conditions, local governments and community organizations must document neighborhood features. Urban planners have relied on neighborhood audits to systematically observe and record features such as neighborhood safety, walkability, etc. However, the physical audit visits are time-consuming and expensive, and Street View Image provides an economical alternative to conducting `virtual` audits. As we discussed earlier in the Background section, Google Street View image is the most widely used street view image service to support this need for conducting observational audits due to its panoramic view, geographical coverage across cities and countries, and user-friendly access made possible via interactive web application. 

Our participants have used these images in purely ethnographic work where the ground-level perspective of SVI allows for understanding and analysis of fine-grained neighborhood characteristics or human behaviors in space. For instance, P3 described sending staff weekly to survey changes to retail spaces on the ground level, currently taking up to 20 person-hours of labor a week using ``a spreadsheet they actually print out and walk around with and check on things''. As an alternative, P3 mentioned recently using street view to check ``if there were manhole covers on the street'' which were covered in snow at the time. However, P3 also added that to their staff, GSV is a tool primarily for when they are ``being lazy'' or need to view something far from where they are. P2 described the use of GSV in understanding urban living as,

\begin{quote}
    [Part of the job of an urban designer is understanding] how do people move through the space. Do you notice that a lot of people sit there? Are there always constantly groups of people sitting around? Or is there always a steady flow of people walking through the space? I would also [pay attention to] cyclists. Are there any cyclists moving through space? Are they locking their bikes anywhere?
\end{quote}

Tools like Google Street View are valuable to many community organizations as they provide simple outputs that capture the story and translate to policy design. The designers are ``able to maybe copy and paste specific data points that were helpful for their work....create screenshots of custom maps that they can they can put in their report'' to tell the story about the neighborhood's needs to their stakeholders. P2 noted that for users with less technical experience in interacting with large datasets, ``the biggest value was really the ability to visualize data and do some customization of maps and graphs''. 

\begin{quote}
    [Community board members] weren't data experts per se, they just have these reports. The biggest challenge that they had is that the [NYC] open data portal seemed overwhelming. They're not comfortable with spreadsheets. They have these reports, but they're often outdated...The audience honestly just want the visualization that were helpful for their work. 
\end{quote}

The lack of tooling limits fine-grained modeling of services and activities availability across different areas in the city, as P1 explained its effect in planning decisions, ``[Because] we're applying the same standard to every city block''. However, characteristic needs depend on neighborhood contexts such as demographics, age, race, economic factors, etc. One way used by the participants to identify these needs is interviews and focus groups, where stakeholders come from the local community ``would talk about how the [city] landscape is for the older adults [for example], what are their needs", as P1 noted. Interactive dashboards or tools for street view images can become particularly helpful to pinpoint a ``particular intersection that is dangerous for older adults)'' [P1],  answer questions about physical and social properties such as ``Are people sitting there or large groups of people gathered? What kind of the demographics of the people are there? Are there a lot of kids in the photo?''[P2] or verify physical locations of streetscape service issues such as ``large potholes, sinkholes...or graffiti...anything that needs to be maintained or repaired [based on public reports made to service hotline]'' [P3]. P2 nicely summarizes the importance of street view imagery and the lack of tooling to connect imagery and analysis together that urban planners can use: ``The reason why a lot of photos are required for city planning or zoning applications is that you can read a lot from photos of places in the city. I just haven't seen yet a lot of tools that make that connection. [Tools that try] to further analyze the data that [urban planners] are reading from these photos.'' 

\subsubsection{Limitations.}
While using a street view image as a way to view a physical location without making a physical visit is a great resource, services like Google Street View can raise issues such as data inconsistency and unreliable coverage, as pointed out by our participants. Google Street View cars typically revisit a street around once or twice a year, often optimizing for times when the street is the least busy. This limits the ability to use their data for understanding aspects of the urban environment that change faster than that or focusing on people and objects moving through the space rather than aspects that are fixed and infrastructural. Image update schedules are determined by factors such as time of the day, weather, amount of traffic on the road, etc., which can introduce temporal bias, a type of data bias arising from changing data behaviors over time. 

P5 uses Google Street View images to characterize the built environment from a computational perspective such as urban greenery, bike and pedestrian counts, and they describe their concern over this lack of temporal diversity, ``[...] so the challenge over there is the Google Street View is not very consistent. Although it's updated every other year, but it's not taken at exactly the same time. We have projects where we need more consistent urban greenery measurements. So when it's summer versus when it's winter is a huge difference.'' Similarly, densely populated, urban areas are updated more regularly compared to rural, lower populated areas,  which can impact the representation reflected in the sampled data and the uncertainty level in the measurements calculated, as outlined by P5:
\begin{quote}
The limitation is that because Google Street View only covers street networks throughout the country. So if there is a street that is only available for biking, and for pedestrians only, then you don't really get much image there. And that could be a problem if you want your Local Climate Zone [urban climate zone classification] prediction to cover the rural areas where the image update is also not very frequent.
\end{quote}

The trade-off between the value and cost of the dataset is a deciding factor for using street view images in urban planning projects. Much of the popularity of street view image comes from the growth in its use for urban analysis and research, which the current cost model no longer supports. Both P5 and P4 mentioned how the change in Google Street View pricing affected the research that started during the free access period and compelled them to move to open-source, economical alternatives. P4 described: 
\begin{quote}
    [...] the value of a data set changes based on the competition [...] if it gets too expensive, people are just going to maybe not care anymore, and then find another way to do it or start collecting data [themselves]. The reason why people get interested is because it's free and low-hanging fruit, but once that paywall gets introduced, it starts to interfere with that scale and seems less useful.'' 
\end{quote}

P5 explained a similar barrier experienced in using proprietary street view images, ``We [were hoping to form] a certain kind of collaboration where we access their data, then use our own algorithm to develop and compare with their own data. [We found out] that we need to purchase the data. That's the point where we were like, ``Oh, maybe that's not what we need.'' These excerpts from participants highlight how despite the ease of use, availability of interactive tools, and standardized data collection have made current street view image datasets a widely recognized data source, quality issues like inconsistent spatial and temporal coverage and availability cost can influence the value and utility of these datasets.

\subsection{Opportunities and Applications for Novel Street view Images}

Our participants shared that of the use cases for street view imagery currently explored in present works (discussed in section \ref{sec:background}), many would benefit from data updated at a more frequent schedule as this allows exploring questions that require data collected at higher frequency. P5 suggested using air quality monitoring as an example that ``projects that try to predict air quality based on Google Street View images if it's updated every minute, will result in a better [fine-grained] air quality model.'' In particular, GSV-like services are more appropriate for studying static attributes that experience slower change over time (i.e. sidewalk width, slope, etc.). On the contrary, existing high-frequency street view data providers offer permissively licensed images collected by crowd-workers and can offer more frequent sampling, but heterogeneity in quality and spatial coverage remains arbitrary in these datasets. P4 described this as a trade-off between quality and quantity, ``Like [you can get] taxis for 24 hours, and it would cover every block. What happens if you just pick 30 taxis at random and [that's kind of] what it would take to at least get to each block once, so, of course, there's some variability in that... so the big trade-off is quality vs. quantity''

Temporal granularity can enable projects where variation occurs due to seasonality (i.e. greenery), or weather (i.e. bike and pedestrian counts). P1 noted the importance of frequent temporal sampling for cases where, ``this block could be fine [you know,] for nine months of the year, but three months of the year it's treacherous. 

\begin{quote}
    You can have a street that [by most definitions] might be quite walkable except two days a week due to the garbage pickup schedule...It could be even more granular as the time of the day, for example, if you live across the street from a school, a high school, or an elementary school, you know, maybe most of the time walking in that community is not a problem. But when the kids are leaving school three o'clock or right before school starts, there's no way that you can get access, so it could even be more granular than you know, a whole day or times of the day.
\end{quote} 

P3 focuses on local business improvement and suggests that traffic counts at different times of the day can help in designing policies for on-foot traffic, ``For instance, I thought that [our peak] counts [for bikes] would be in morning and evening. In reality, was actually lunchtime and evening, and what we realized pretty quickly was that it was  largely commercial cyclists at lunchtime, making deliveries.'' In a similar spirit, P5 describes potential opportunities for higher frequency data in understanding micro-mobility, ``a problem that hasn't been solved so far, you probably have frequently seen drivers just double-park there,...and loading and unloading, and a lot of places are trying to get rid of on-street parking...which frees up the curb spaces saying, "This space is for loading and unloading only."  
\begin{quote}
    How do you define how long people park there? What would be considered as loading and unloading? Why are people sitting there for 30 minutes just for loading or unloading? That's very hard to decide, and also, like, all those enforcement would benefit from the images that are updated in real-time, so the images will be also very valuable.
\end{quote} 

Even understanding the quality of built environment and space over time can be informative as P2 outlined, ``the quality of the space over time [such as] is the bench still in good condition, has it broken down in any way, or the space has just degraded in quality or you maybe some [building] textures are crumbling...or vandalism of some sort ...maybe trees fell down in a storm.'' Overall, a recurrent theme appears in our participant's perspectives that street view image services that are available at different time scales hold opportunities for many new applications which are not possible with services that are currently available and accessible to them.

\subsection{Obstacles Experienced and Tools to Mitigate Them}

Many promising applications in the urban planning domain can be made possible with emerging street view image datasets. However, our participants delineate that regardless of the data services one uses, there are some fundamental barriers that they encounter when utilizing SVI in their analysis process. One fundamental barrier our participants shared is assessing the cost of using a new source of data, to make budgetary decisions. Deciding to use a data product is intimately related to its cost. The cost of access and cost of use can be prohibitive to organizations, particularly when researchers are uncertain if they will be able to extract relevant outputs. P2 describes this assessment process as, 

\begin{quote}
    There were a few times when there were proprietary data sets that we could use. However, they were either too expensive for a city budget and you have to go through a whole procurement process that can open up a can of worms... At the end of the day, we kind of scale back the scope of the kind of analysis we're trying to do with just open data sets by trying to assess how much value we would really get for this proprietary data. Is it really worth that cost whether it is financial cost or the labor costs of doing this procurement process?
\end{quote}

Conflicts with data-sharing policies and privacy concerns can also influence the burden of procuring data from the providers and their use cases. P2 shared their frustration on this topic: ``We can't just rely on open data because it's a good starting point but it's not always accurate for things you wouldn't think of...items that change very frequently.'' The value of information extracted from the data depends on the quality, and without an interactive, exploratory tool that allows focusing on certain geography or demographics, organizations with limited budgets for labor and technical infrastructure rely on open-source data. Government resources are limited and data procurement involved enough work that it was often better to ``scale back the scope of the work``. P2 describes how time and resource constraint plays a role in this:
\begin{quote}
[The question is] how much time would be held up because we have to get the right permission to be able to get this data in the first place. I think the biggest part of that decision making is just \textit{do we have the time?} And then the second part is, \textit{do we really have the right people to be able to ultimately utilize and do the proper analysis} that this data would potentially give us. Oftentimes the answer was 'no'.
\end{quote}

On the other hand, the restrictions on data usage, resulting from pricing policy changes, can affect the degree to which academics could rely on GSV and turn to other data sources, as outlined before. This also creates a data inequity as access to proprietary data becomes more available to institutions and organizations with larger budgets. 
     
Another obstacle, that is directly associated with cost is the ability to query and filter data based on analysis needs. While one can download images in bulk using Application Programming Interfaces (APIs), storing and managing the data becomes complicated once outside of the web interface. Our participants shared how that impacts their ability to extract insight from the data and its practicality across areas, ``if I do walkability score, for example, it is derived on a whole neighborhood rather than [like] a specific street. [or] If I'm using demographic data, [it captures] a whole neighborhood.. or a whole part of the city, rather than a specific street. If I'm interested in trying to understand the walkability for seniors or people with disabilities, it would be helpful to know if that data is available, how limited it is, is it that we only have that data available for [selected areas such as] parts of lower Manhattan. Finding data for very specific groups of people is helpful.'' 

The assessment to procure relevant data one needs depends on the specific research goal. 
Frequently sampled, spatially sparse,  data collected at a regular interval, is useful when monitoring physical assets (i.e. trees, street signs) that may change more rapidly. But research processes that involve computational analytics often need to use the ability to filter data based on their value to answer research question. When a large number of images are taken in the dark, blurry, poor resolution, this affects the information value of the image where computer vision based techniques are applied to classify sky view, greenery, or pedestrians. P4 describes the issue using facial recognition as an example, ``The contrasts and the shading of the photos, the resolution, the blur can make some of those [AI] algorithms fail on their face right.'' One analog to this varying quality of SVI that P4 pointed out is the challenges experienced in analyzing satellite imagery and removing quality issues such as cloud coverage and motion blurs, which are not yet available for SVI. As a result, researchers and practitioners rely on collaborative projects to resolve budget constraints for data procurement. ``but we are lucky
it's like a multi-nation collaboration project.
We're standing on the giant's shoulder. The first million images were expensive,
but the additional million images were much cheaper. So, which gives them less motivation to say,
\textit{I want to query the data and then do this and do that.}...Basically they [X institution] are rich enough, say, \textit{I want all the images.}", P5 adds.

One way to adjust the data cost apportioned by its quality and value is through exploratory interaction. The lack of available tools and interactive dashboards has limited the ways practitioners can explore the images to assess various quality issues such as temporal freshness, relevancy, redundancy, and sparsity. According to P3, `we're still pretty dependent on the dashboards that we actually pay for...The actual counts that we are mostly saving via pretty low-tech, so it's not too exciting...for more intelligent planning decisions [that can be possible] with more frequently updated data.'' While the dashboards allow data exploration, making financial decisions that involve filtering data based on quality attributes needs a mechanism that allows researchers to rank data segments (a ZIP area or a day in the week) in comparison to other segments. A ranking mechanism to assess the quality concerning other data segments can help incentivize data collection, and encourage crowd-participation through gamified data capture. Such systematic data collection and quality investigation can also reveal any frequency bias that may occur due to certain areas being sampled more or less frequently, as P5 pointed out:
\begin{quote}
I don't think [Google Street View] it's the best data to characterize where people are given a certain time of the day because it's biased [from data collection] towards the time when there aren't many people in the city... Or if you are on a biker, you will try to avoid roads
with a lot of vehicles. That's where those cameras are scanning more frequently. So say, I want to count how many people on this street actually use bikes, that is a biased model, right? So, maybe the demand will be much higher in streets where you don't even have your camera scanning over there because it's bike-only or pedestrian-only. And there's no way you can validate it because you don't have ground truth.
\end{quote}

Overall, our interview study with experts illustrates that novel street view images have opportunities to contribute to research endeavors that present data sources do not provide. However, users of these data need tools and frameworks that will help them assess the value of the data concerning its quality and cost. Such assessment can help data users compare multiple data sources that are available in the market and procure the data that best serves their needs.









\section{Design Solution}\label{design}

\begin{figure}
\centering
    \includegraphics[width=.90\linewidth]{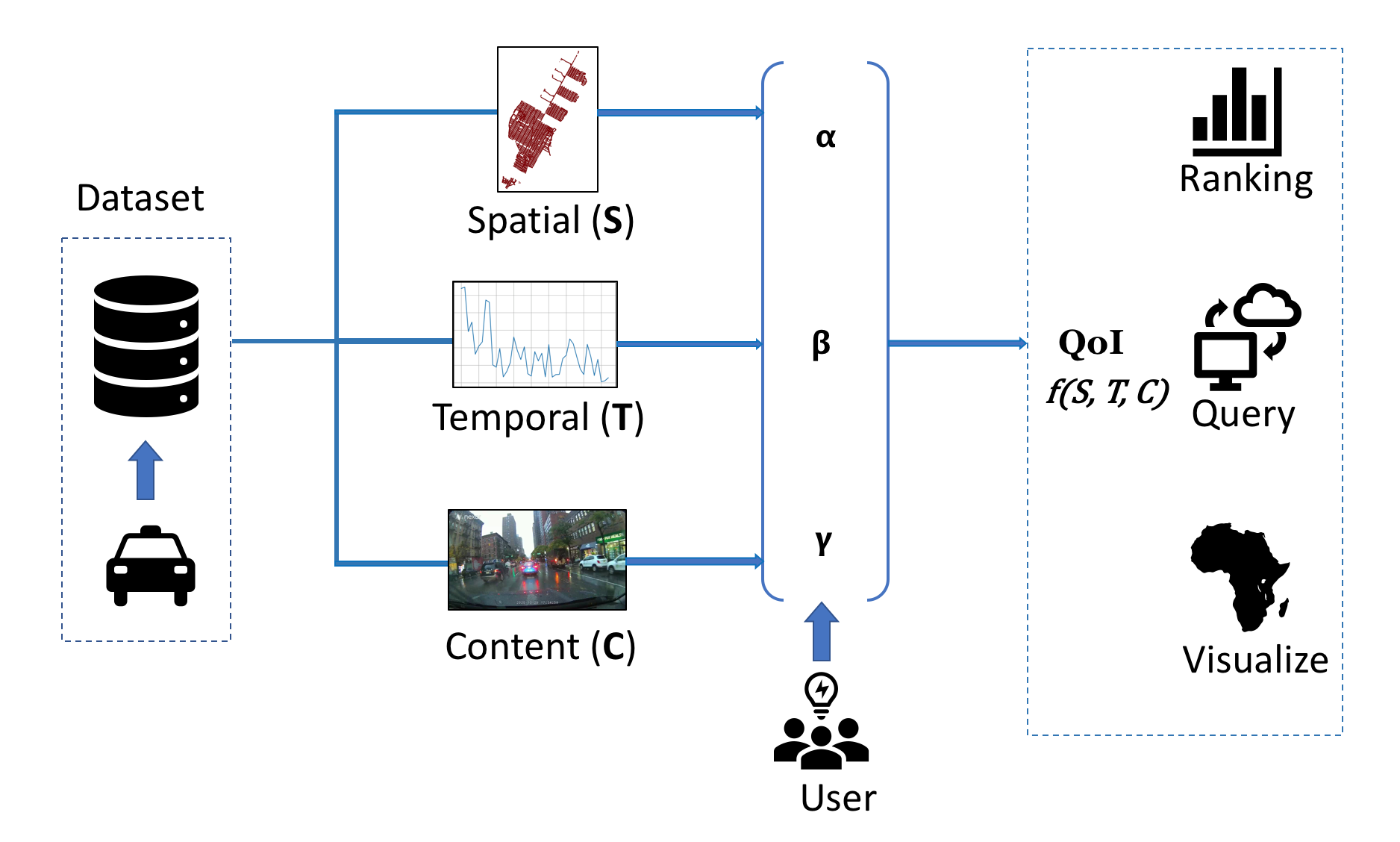}
    \caption{Overview of the proposed Quality of Information framework based on spatial, temporal, and content attribute dimension. Such framework can be used to rank, query, and visualize information relevant to specific use cases.}
\label{fig:Framework_Overview}
\vspace{-10pt}
\end{figure}

Building on the insights from our formative interview study to glean uses, challenges, and opportunities of novel street view images with high fidelity, we designed a framework that can be used to evaluate the quality of the data with an aim to mitigate the obstacles experienced by practitioners in using them. Understanding the interdependent relationship between the value and quality of the information in street view images to estimate its perceived usefulness is the primary step in the stakeholder's workflow, we focus on designing the framework for estimating the Quality of information of SVI. Specifically, based on participants' views on opportunities and challenges of utilizing street view imagery in urban analysis, we have identified three attributes that can address the quality issues of spatial sparsity, temporal frequency, and content relevance. Researchers and practitioners can use this framework
to evaluate data quality along each attribute dimension and determine its value to address specific research questions relevant to a wide range of applications or use cases.

\begin{figure*}[t!]
\centering
    \includegraphics[width=.235\textwidth]{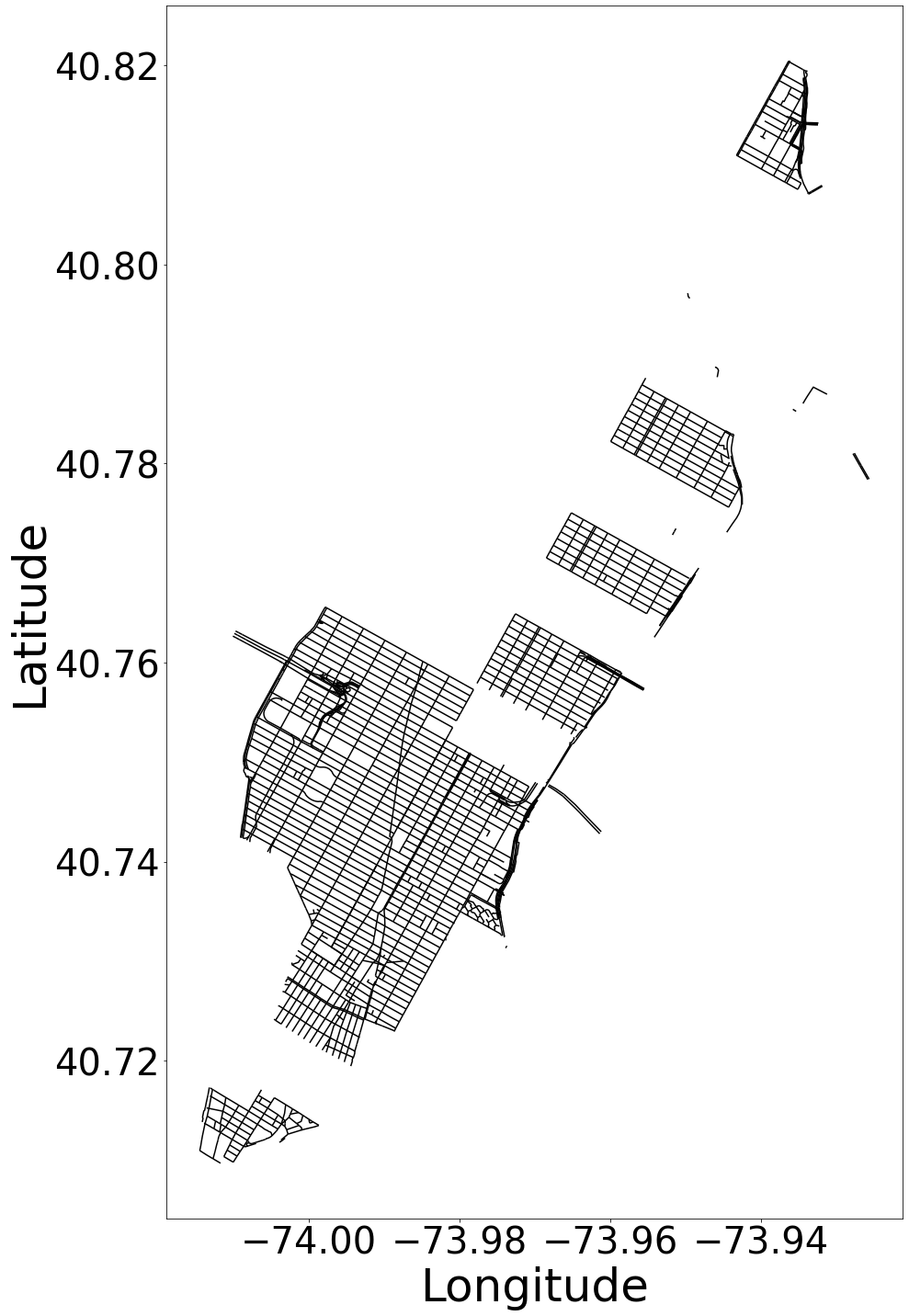}\hfill
    \includegraphics[width=.239\textwidth]{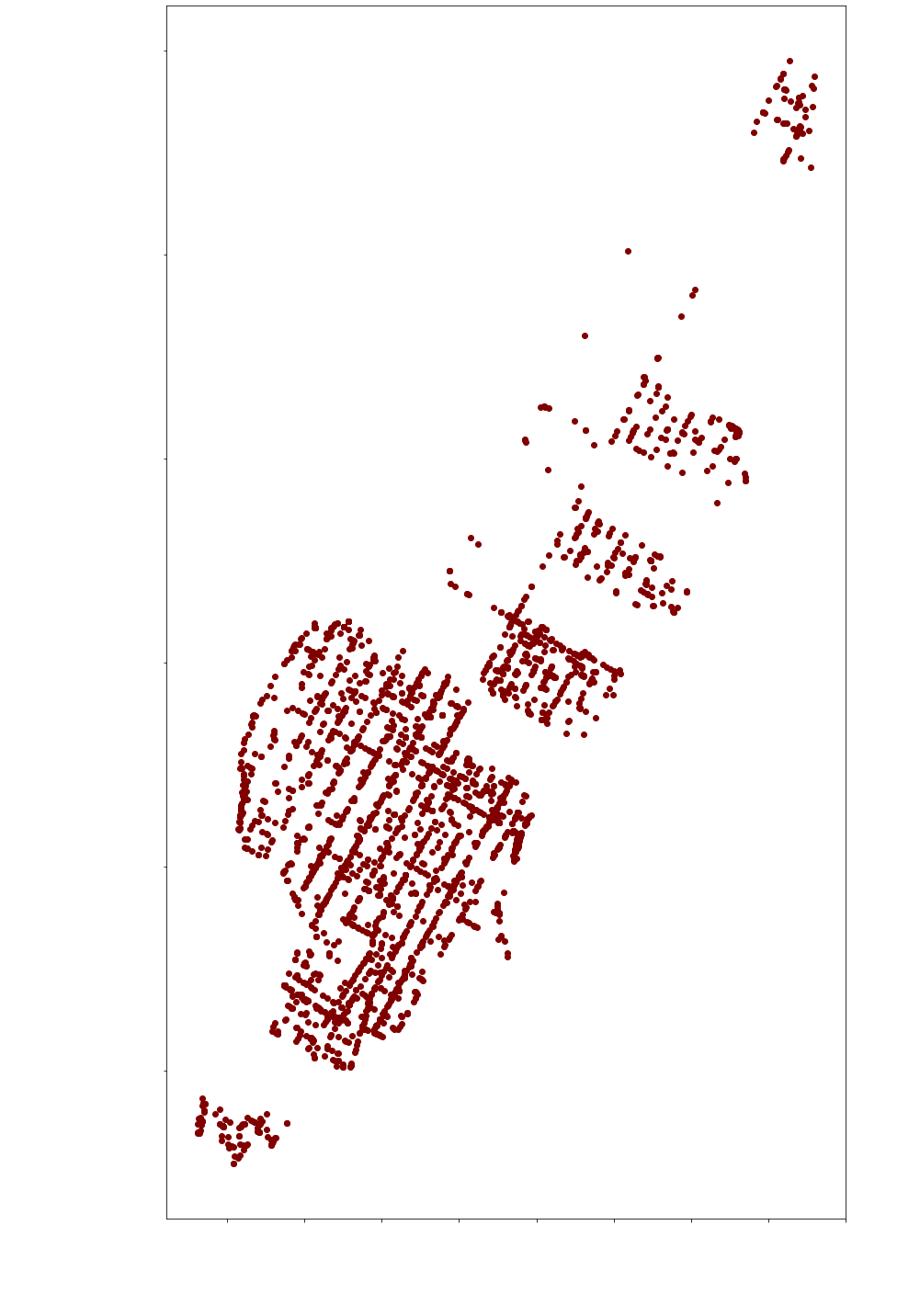}\hfill
    \includegraphics[width=.239\textwidth]{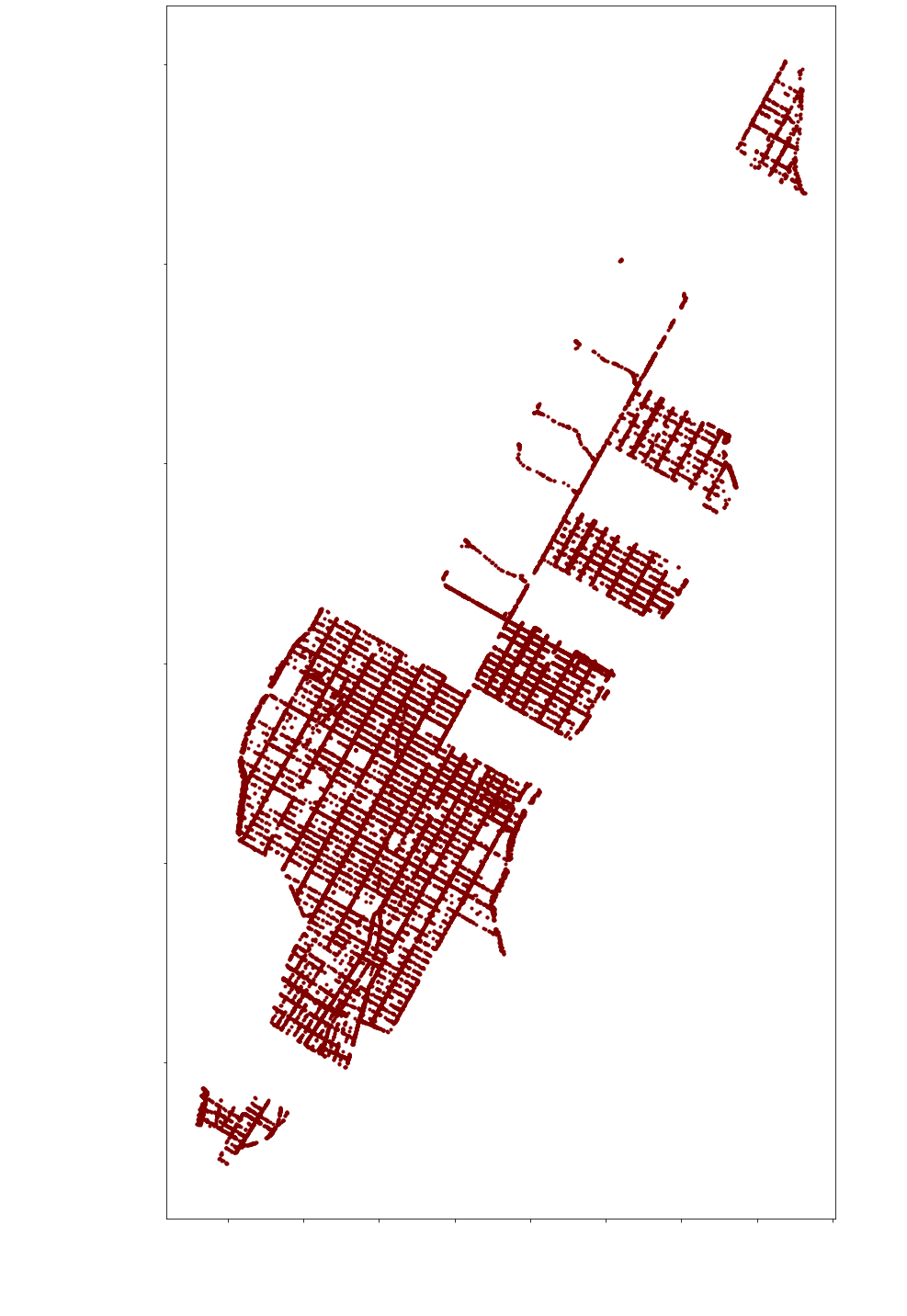}\hfill
    \includegraphics[width=.239\textwidth]{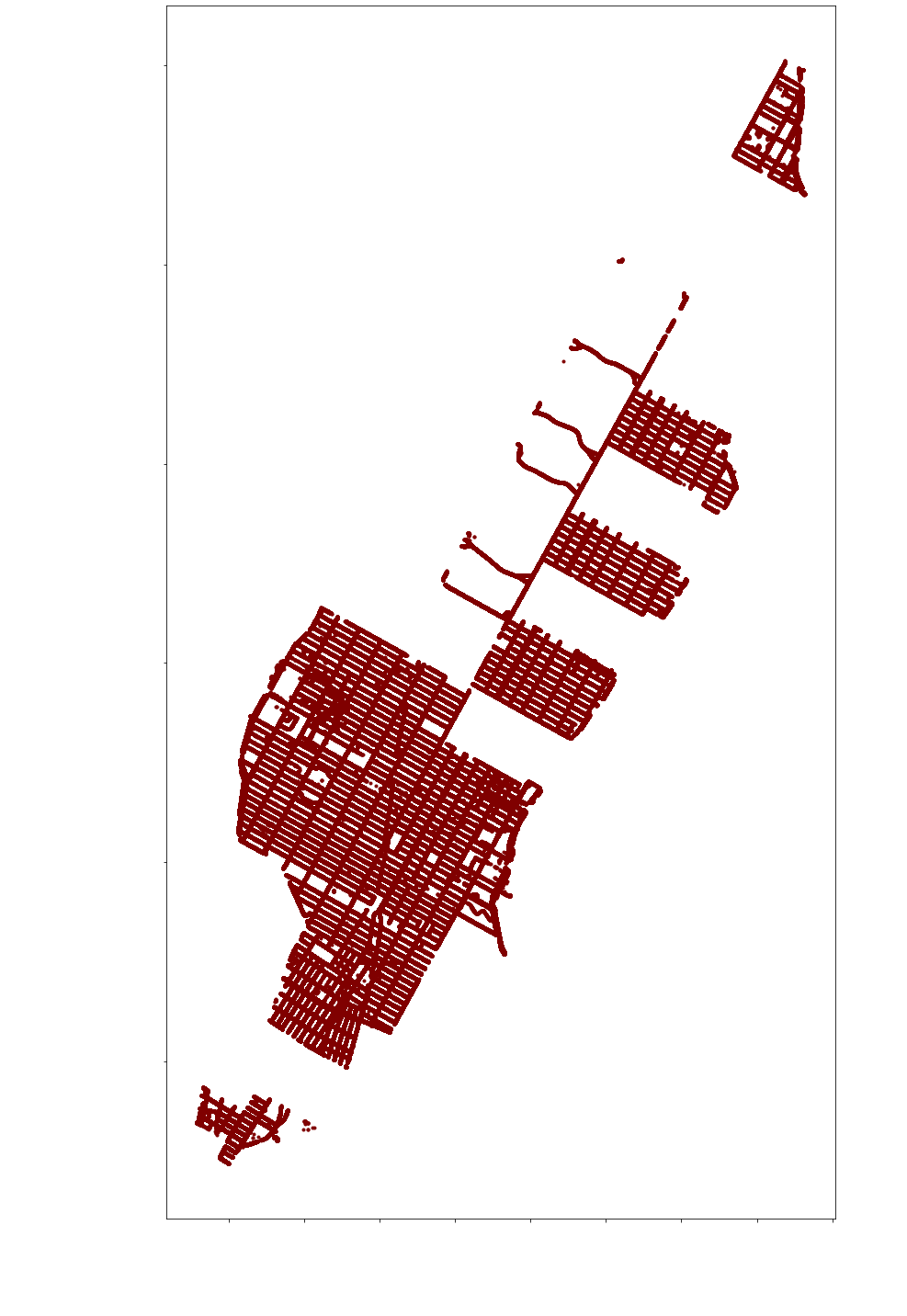}\hfill
    \caption{Spatial data distribution over different time period, a) street network of 14 zip code areas, b) street view data distribution for a single hour, c) a single day (24 hrs), and d) entire data collection period (46 days).}
\label{fig: data distribution}
\vspace{-10pt}
\end{figure*}

\subsection{Attributes for Large Scale Street View Image Quality Framework}

We conceptualized the attributes in our framework based on the current barriers experienced by users as illustrated in our participants' views. Specifically, we focus on measuring the quality of a data segment, query data based on research needs, and rank data segments based on their research needs. The large volume of data generated by vehicular networks is challenging due to the processing and storage requirements. These issues may be mitigated by querying and filtering data points below certain information quality (e.g. filtering images in dark, occlusion, blurriness, etc. conditions). Images may be of sufficient quality for some applications while needing a higher quality score for others. Collecting more information or maximizing coverage does not guarantee improved data quality. For example, air quality changes slowly, so a higher sampling frequency may not add more value to the data in regular environmental sensing applications. Evaluating the value and relevancy of data is essential to justify the monetary and labor investment urban professionals need to sanction that can help in financial decisions for data procurement to incorporate such large-scale datasets into their analysis workflow. Building on such a quality assessment framework, data-driven mechanisms can be formulated to localize data segments with low information quality and incentive mechanisms can be designed to improve the quality of captured imagery~\cite{10.1145/2888398}.

\subsubsection{Spatial Quality.}
Most street view image services are considered as a `digital twin' of the street scenes in the real world. Maximum spatial coverage is thus an attractive property for many applications where information is extracted from images and their distribution is mapped across geo-spatial areas to gain urban-scale insights. A real estate developer may be interested in estimating car counts and prices for modeling housing affordability in a single block, whereas non-profit organizations studying older adults' mobility would want to map sidewalk conditions and slopes across an entire city. Most researchers working on street view image-based urban computing, design, and virtual auditing want a dataset that captures real-world spatial observation as close as the true distribution. This allows them to minimize representation bias, which arises from how data is sampled, making decisions based on observations that are not sufficiently comprehensive to represent the real world.

\subsubsection{Temporal Quality.} Many urban experts are interested in application areas that require revisiting streets to update imagery at high temporal frequency. This is particularly useful for understanding urban aspects that change fast or move through space needing the dataset to be updated at different temporal granularity such as every day or at different times of the day. A researcher may be interested in urban mobility to map times of the day with older adults' difficulty navigating the city or curbside parking availability in the city. Google Street View images are limited for such purposes as the collection procedure optimizes for times when the street is the least busy and is updated once every 6 to 12 months (in urban areas). Being able to query crowd-sourced images for such temporal insights is valuable as they provide much higher temporal resolution, especially in areas frequented by many people.

\subsubsection{Content Quality.} The quality of crowd-sourced street view image content can vary due to device heterogeneity, variance in orientation, image conditions (e.g. occlusions, poor lighting), etc. Images contain `general-purpose' information with many use cases. Regardless of whether a practitioner is interested in spatial or temporal insights, extracting these insights through expert manual auditing or advanced image processing techniques (detection, segmentation, etc.) automatically, depends on sufficient quality in image content and is essential for its usability in analysis. A blurry image making it difficult to identify street objects can be analogous to a sensor reporting zero readings in an air quality monitoring application. As large-scale street view images are collected in an uncontrolled setting and used in many application areas, a quality evaluation approach scalable to many use cases can aid in need-based filtering and optimize data cost.
    



Based on these attributes identified, in the next section, we describe a framework for computing information quality along each attribute dimension using a street view image dataset. While we use urban-scale image data as an example dataset in developing our framework, we believe the framework for spatial and temporal attributes can be scaled beyond images to other urban-scale sensing data.

\subsection{Framework Design}

We now describe the design principles of the quality of information framework for street view images. Specifically, we outline the methodology to measure three quality attributes related to spatial, temporal, and content information in a mobile sensor network.

\textbf{Spatial Coverage.} Consider a two-dimensional geo-spatial area $A$ for which we are interested in evaluating spatial information quality. A sensor (in this case, a vehicle or pedestrian) senses and detects events about a point of interest when the point is covered by at least one sensor measurement. Unlike coverage in a geo-spatial region where points of interest can fall anywhere within the region, data collected on the streets would only cover points along the street network (This is a limitation for data sampled by street view images does not cover areas where vehicles cannot enter, i.e. gated communities, bike only streets, etc.)\\ 

If each street in that area is $r$ and the complete network of streets is $R$, where $r$ is a subset of $R$ ($r \subset R$). Thus, following~\cite{10.1145/1062689.1062728}, the area coverage of a sensor network during a specific time interval $[0, t]$ is the fraction of the street network covered by at least one sensor within that time frame. However, a mobile sensor network covers a fraction of the area simultaneously at any given time and covers all the points of interest over long intervals. See Figure~\ref{fig: data distribution}), where spatial data distribution of street networks moves closer to the true distribution over time by covering missing holes through repeated measurements. Considering maximum coverage as the closest to the reference coverage scenario, we can estimate the quality of spatial coverage obtained over a given time interval (a data segment between two different days or times of the day) as the distance between reference spatial distribution and true distribution observed in the data. 

If we consider a multivariate uniform distribution $P$ for reference and observed data as a multivariate normal distribution $Q$, the Spatial quality $S$ is measured as,
\begin{equation}
    S = d (P, Q)\label{eq:1}
\end{equation}
where $d$ is a distance measure for probability distributions. While there are many distribution distances available in the literature, in practice, we consider two measures: 1)  Wasserstein distance, a measure describing the minimum average distance required to cover so that distribution $Q$ moves to distribution $P$ to become identical, and 2) Jensen-Shanon divergence, a measure of the similarity between distributions $P$ and $Q$. While there are other distribution distance measures such as Kullback-Leibler (KL) divergence, Maximum Mean Discrepancy (MMD), etc., we choose the aforementioned two measures because of their symmetricity property and fast computation based on our empirical results. 


\textbf{Temporal Frequency.} Assuming mobile sensors move independently of each other, available sensors at any given time cover a fraction of the points of interest which over time leads to covering all points. However, considering street segments in an area create an interconnected network with several alternate paths to traverse to a specific location, points are likely to be revisited during a time interval depending on sensor density. Thus, a fraction of points along the street network have a high probability of being updated at a higher temporal frequency, even multiple times in a day.

We consider a specific point in the dataset, which for a specific temporal interval $[0, t]$, is sampled at $n$ random times $(t_1, t_2,..., t_n)$, where $n>1$. Considering there exists a total of $N$ such points, the dataset queried for such $N$ points thus provides temporally rich information for applications that change at a faster rate and need regular monitoring. If a point is sampled $n$ times at random time intervals in a sensor stream, we can compute the sampling rate by considering sample intervals of consecutive samples and find the dominant sampling rate ($u$) for the considered area. Using the spatial displacement formula ($s = vt$) as an analogy, temporal coverage of the point, $T$ can be measured as $T = 1/u * n$. Summing over $N$ points, the total temporal quality of the area is,


\begin{equation}\label{eq:2}
  T =   {\sum}_{n = 1}^{N} (1/u) * n 
\end{equation}
 where $1/u$ represents temporal resolution per sample. We can sum temporal quality over the entire time period of the dataset to evaluate temporal quality for a time duration longer than $t$.

\textbf{Content Quality.} Advanced image processing techniques used for feature extraction often perform poorly on low-quality images such as blurry, occluded images. This is reflected in machine learning algorithms' low confidence in well-known computer vision problems such as object detection. Human inspection may be more informative in these cases as experts can apply disciplinary knowledge to evaluate the relevance of the image content to the expected analysis. However, this does not scale well when the dataset contains millions of images. Images with poor lighting conditions can provide lower analytical value due to the difficulty of interpreting image content.

As object detection algorithms are frequently used to extract semantic information from street scenes captured in street view images \cite{wei2019city, wu2021learning}, confidence scores for objects detected in an image can be considered as a proxy for image quality. As object detection and image interpretation, in general, is a hard problem for images taken in low light conditions, we consider a subset of image samples collected in a geographic area $A$ over a time interval $[0, t]$ filtered based on the average brightness level of the images. Considering $x$ number of objects detected in a specific image, the average confidence score per object $c$ is considered as content quality for the image. If $N$
number of images are detected in images collected for that area in that time interval, total image content quality ($C$) can be measured as,

\begin{equation}\label{eq:3}
    C = {\sum}_{n = 1}^{N} c
\end{equation}

For implementation simplicity, we use YOLOv5 \cite{redmon2016you}, an object detection method frequently used in urban computing for detecting objects, and obtained average detection confidence as a measure for image quality in downstream applications. Note that YOLOv5 is used here as an exemplar analytic process for extracting insights from street view images. Framework users can replace this measurement with other automated image processing techniques (semantic segmentation, scene description, etc.) and develop appropriate uncertainty metrics to assess content quality.

\subsection{Putting it all together into a unified quality metric}\label{customize_QoI}

While we are interested in delineating specific attributes in street view images to empower users to focus on their desired quality dimension, a unified quality of information metric can tie all three attribute dimensions into a single metric to aid in analyzing data quality holistically. Such a metric can assess a data segment collected in a specific geographic region with respect to other regions based on spatial, temporal, and content information, rank them, and query data based on certain quality standards. 

The quality of information for a geographic region $A$ in a specific time interval $[0, t]$ can be formulated as $Q$ = $f(S, T, C)$, where, $S$, $T$, and $C$ are the quality of information along the three attribute dimensions \textbf{S}patial, \textbf{T}emporal, and \textbf{C}ontent. A researcher interested in all three dimensions may put equal importance on all three dimensions, whereas others may emphasize spatial or temporal dimensions for their research needs. While we acknowledge that an application-specific QoI framework will be deemed most effective, we prioritize generalizability in our framework design so that researchers can customize the framework to their needs and data. Thus, the unified quality of information model can be interpreted as a linear combination of the individual attributes,
\begin{equation}
    Q = \alpha \cdot S + \beta \cdot T + \gamma \cdot C
\end{equation}
where, $\alpha$, $\beta$, and $\gamma$ are coefficients of each independent quality attribute for the weighting mechanism. 

\textbf{The choice of coefficients.} Note that the QoI is a linear combination of the individual attributes and the co-efficient will be dependent on the preference of the user. The preference, if made through rational decision-making, will also depend on the researcher's objective and constraints. Towards this end, we can pose this as an optimization problem, where researchers need to maximize value while minimizing cost. Such constrained optimization for decision-making can be helpful to incorporate the budget constraint, objective function, and quality variables into a mathematical analysis model, and solve through approaches like linear programming.

As the constraints and objective function will vary based on applications, we present the framework in this work with user-defined co-efficient values selected based on disciplinary heuristics. Learning algorithms, which select co-efficient of importance based on what has worked well in the past, can be an alternative consideration to include objectivity. However, our approach here is informed by works in decision sciences on competitive priorities~\cite{ward1998competitive} and prior works comparing human heuristics to algorithmic decisions that suggest heuristics performance is comparable~\cite{pmlr-v58-buckmann17a}.

Inspired by works in psycho-physical scaling for comparative judgment analysis under decision heuristics \cite{saaty1988analytic}, the value for attribute coefficients in our framework ranges between 1 through 5 which resembles heuristic scales a decision maker may rely on. A co-efficient of 1 for all three attributes considers all of them equally important with a direct correlation to overall QoI while assigning 0 to one of them disregards it and evaluates quality along the remaining two. The advantage of using numeric scale ratings as coefficients is these ratings can directly be applied as weight factors for a particular quality attribute. An overview of the complete framework is presented in Figure~\ref{fig:Framework_Overview}. 



\section{Results and Evaluation}

Based on the framework described in the previous section, we now present a case study examining how the framework can be used. To measure the quality of information for street view images in this section, we used a street view dataset collected from New York City as an exemplar. While this means our data contains images taken mostly in urban areas, the approach can be scaled to rural or suburban areas as long as there exists a street network for cars that are sampled in the data. We present the results of measuring QoI to evaluate how the proposed framework can be used to serve the needs and address some of the challenges experienced by expert users outlined in section~\ref{sec:needfinding}.

\begin{figure*}[t!]
  
 \includegraphics[width=0.32\linewidth]{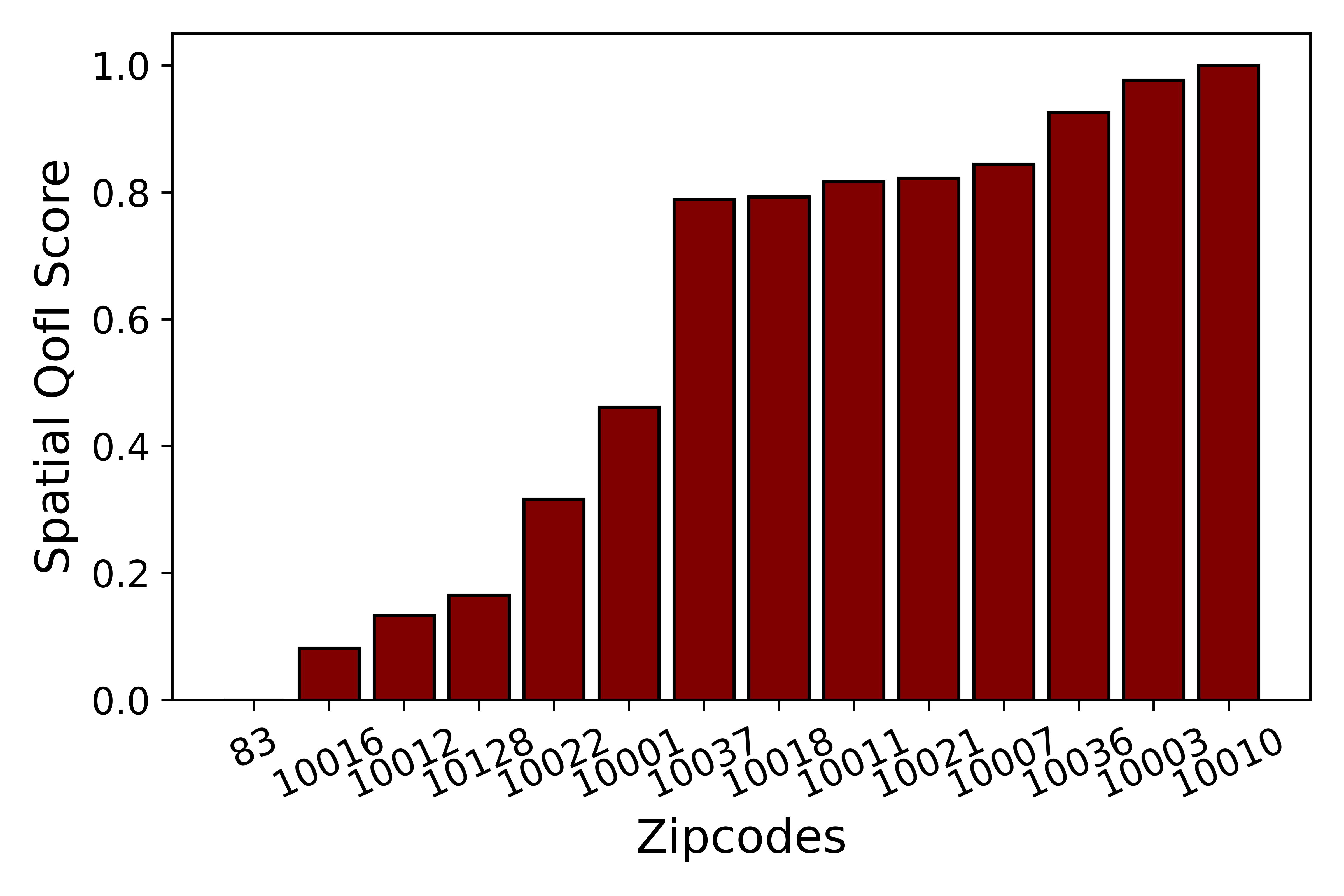}
  \includegraphics[width=0.32\linewidth]{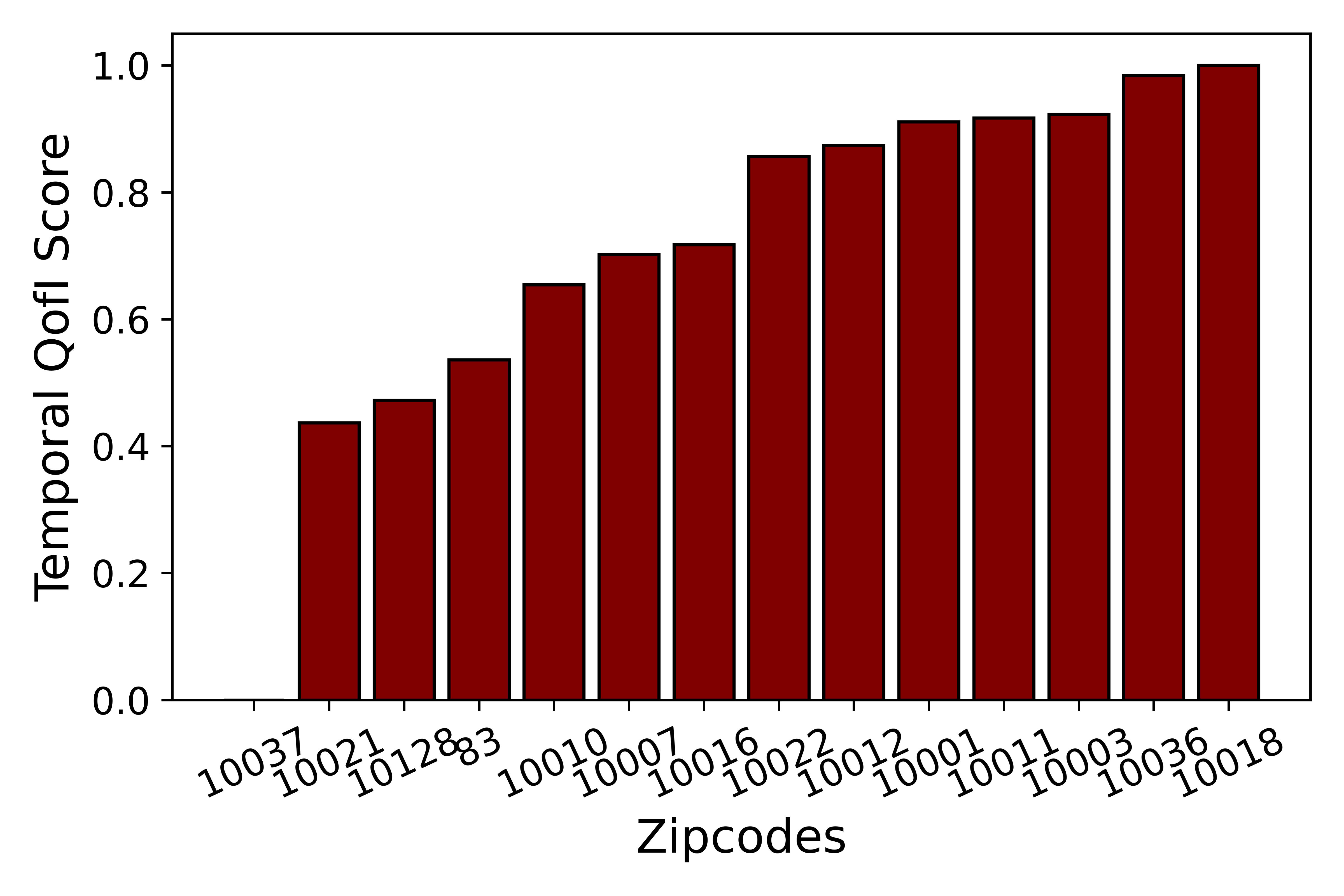}
  \includegraphics[width=0.32\linewidth]{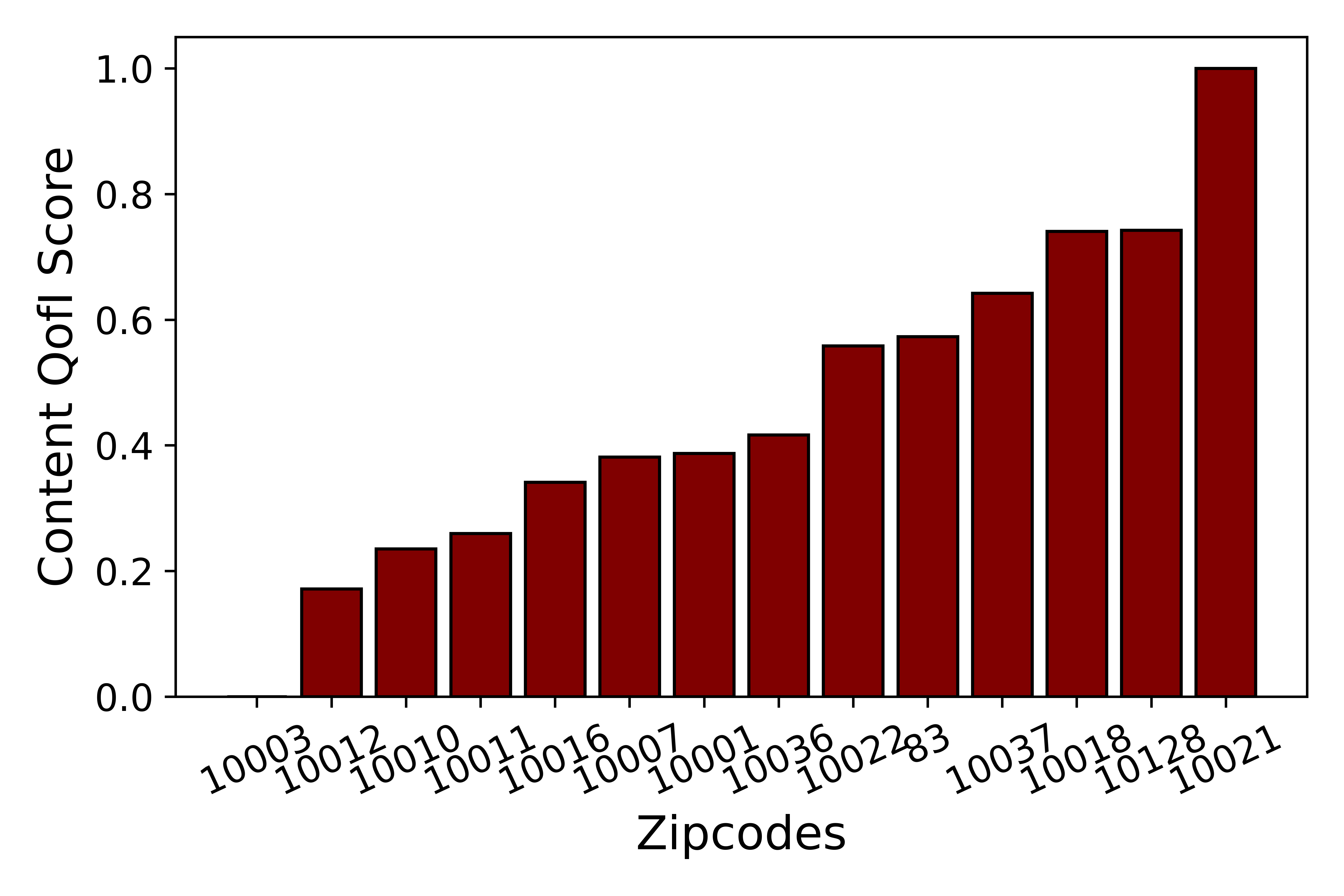}
  \caption{Ranking of 14 selected Zipcode areas in New York City based on Quality of a) Spatial, b) Temporal, and c) Content Information. Such ranking can be used identify areas with high and low QoI along different quality dimensions and devise data collection strategy for improving information quality in low QoI areas.}
  \label{fig:spatial temporal content QoI}
  \vspace{-3mm}
\end{figure*}

\subsection{Dataset}\label{dataset} We use a dataset collected by Nexar, a company providing real-time road data collected from dashcams on taxi and ride-sharing vehicles. 
As of July 2022, when the data was obtained, Nexar claims to host over 1.5 billion miles of data 
that users can interact with on their CityStream platform. Images are processed and objects are detected for mapping work zones, traffic, and road signs.%

We established a data exchange relationship and through this license to Nexar imagery, we obtained access to around 20 million images (
collected from October 5, 2020, to November 15, 2020. The data covers 196 Zipcode areas in New York across all 5 boroughs. Each image has its metadata that contains longitude, latitude, and UNIX timestamp. 

\subsection{Pre-processing}

To lower the processing time and computational cost, we limit our analysis to a subset of the Zipcode areas. We randomly selected 14 zip code areas (mostly from Manhattan and adjacent Queens Borough areas) and the subset of the images that fall within this geographic region. We use NYC Open Data~\cite{nycOpenData} and NYC Department of City Planning~\cite{nycPlanning} to access the maps of zip code areas and the street networks respectively. We binned each image under the corresponding Zipcode area it was captured in which provides us with 14 geospatial segments of the dataset with 2,782,870 images. 

For this case study, we used Wasserstein Distance~\cite{villani2009wasserstein} to measure the distance between reference and observed spatial distribution for its non-negative property that is helpful to interpret by the user. For each Zipcode area, we further binned them for each day of the 46 days captured by the data subset. Information quality along each attribute is independently computed for data captured within a single day. Summing them over all days using the trapezoidal rule to integrate over each time interval (in this case, day), we get the information quality measured over the entire period.  We use the GeoPandas library to convert geo-coordinates (latitude and longitude pairs) to geometric points to visualize on the maps. We use the Shapely library available with parallel computing (CUDA) to determine the number of observations for each point in the spatial region covered by street segments. We then calculate the temporal frequency for each point which is in equation~\ref{eq:2}.  For object detection, we use YOLOv5~\cite{redmon2016you} to detect common objects found in street scenes such as vehicles, bikes, traffic signs, etc. We utilized available Python libraries such as SciPy and PyTorch to implement our framework.

\subsection{Ranking based on Quality of Information}
The findings from the formative interview study revealed that users often need to assess the value they obtain from the data to compare different data sources. We design our framework to help dataset users evaluate information quality for different segments of the data and decide its relative value in their research goal. Using the QoI measures we formulated in section~\ref{design}, we can now rank these geo-spatial segments based on their quality attributes and present them to the user, which we show in Figure~\ref{fig:spatial temporal content QoI}.

\textbf{Spatial Quality.} In Figure~\ref{fig:spatial temporal content QoI}(a), we have ranked the 14 Zipcode areas based on their spatial quality present in the dataset. Each bar represents the distance between the reference distribution (with maximum coverage) and observed distribution from the dataset. The height of the bar represents the quality score corresponding to each Zipcode area. The score ranges between 0.0 and 1.0 with 1.0 representing the highest score.  

From the right, Area 10010 ranks the best in terms of spatial quality, with 10003 and 10036 scoring similarly. 
On the other hand, Zipcode area 00083 provides the lowest spatial information quality with 10016 and 10012 being very close with a score close to 0.2. For researchers interested in improving the spatial quality of a particular data segment (i.e. a spatial area), this can help in making a decision to focus their budget on Zipcode areas 00083, 10016, and 10012 to collect more data. Alternatively, they can use this information to prioritize procuring data segments of other areas when the QoI score meets the certain application-specific threshold. Our exploration of the dataset reveals that areas that contain missing points (`coverage hole') along the street network resulted in lower spatial quality for them, compared to areas where a large percentage of street segments are covered by the observed distribution.

\begin{figure}
  \includegraphics[width=0.80\linewidth]{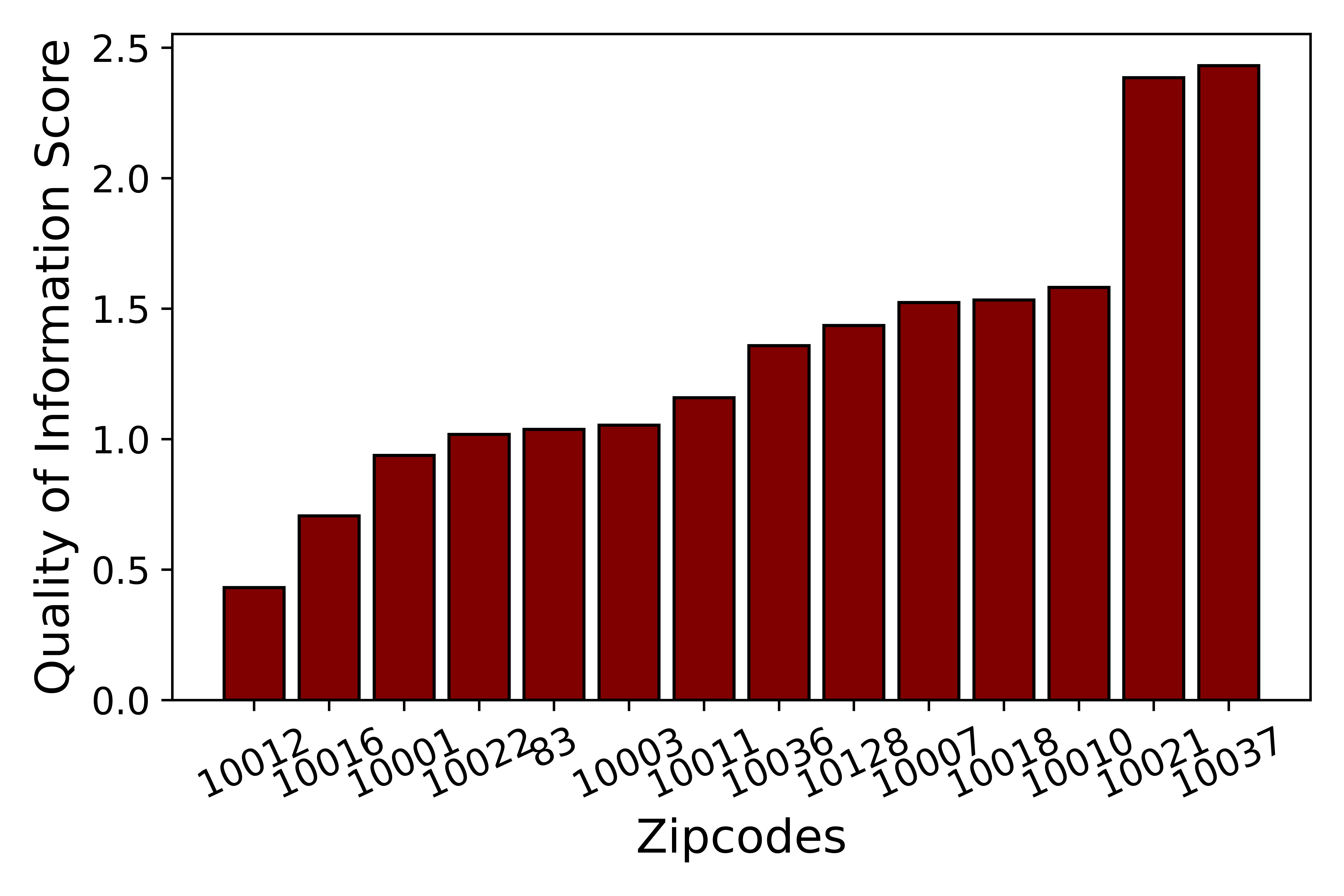}
  \caption{Final ranking order of 14 selected Zipcode areas by composite ranking obtained from individual Quality of Information score from spatial, temporal, and content dimensions. Users can weight individual QoI attributes according to their needs and rank areas based on overall data quality composed of three quality attributes.}
  \label{fig:composite_ranking}
  \vspace{-20pt}
\end{figure}

\textbf{Temporal.} In Figure~\ref{fig:spatial temporal content QoI}(b), each bar represents the temporal coverage score based on the sampling rate and update frequency of each point revisited at least twice. The height of the bar represents the quality score corresponding to the Zipcode area ranging between 0.0 and 1.0 with 1.0 representing the highest. 10018 ranks the best in terms of temporal quality, with 10036 and 10003 scoring similar. On the other hand, Zipcode area 10037 scored lowest providing the lowest information quality with 10021, 10128, and 00083 being very close with a score close to 0.5. We observe that these areas are sampled in a temporally sparse manner (most points are sampled only once) or the update frequency is low during the period, making them less useful for applications that require images updated at high temporal frequency. 

\textbf{Content.} We present the results in Figure~\ref{fig:spatial temporal content QoI}(c), each bar represents the content quality score based on the average confidence score of the street object detected in the images. The height of the bar represents the quality score corresponding to the Zipcode area ranging between 0.0 and 1.0 with 1.0 representing the highest. In this dimension, 10021 ranks the best in terms of image content quality by a large margin from other nearby areas. Next are 10018 and 10128 areas, ranking almost similar. 10003 scores lowest with 10012 and 10010 right before with a score around 0.2. Our manual inspection of the raw images from low QoI days reveals that the low-scoring areas have issues of low light, light scattering due to rain, incorrect mounting of dashboard camera, unclean windshield, etc. which makes detecting image objects for the YOLO algorithm harder. Note that any image processing application that may rely on extracting semantic objects from images or generating image descriptions, will experience such hindrance due to low-quality images.

\textbf{Combined Ranking with Linear Combination.} Through a linear combination of the individual attributes, we can now obtain a unified ranking using all three dimensions. 
For developers and researchers interested in filtering and querying data by a certain quality threshold, such ranking can be useful as it allows them to not only focus on the areas with sufficient quality but also identify areas to improve the value of the overall dataset. To select the coefficients, we use numeric ratings inspired by 5-point Likert scale ratings to measure the importance of the quality attributes. In particular, a rating from 5 to 1 denotes 5 levels of importance: Very Important, Important, Moderately Important, Slightly Important, Not Important.

We present the composite ranking order in Figure~\ref{fig:composite_ranking} where we use a value [1, 1, 1] for $\alpha$, $\beta$ and $\gamma$. 10021 and 10037 have the highest information quality by a large margin whereas 10012 area reports a score that is approximately one-fifth of that of 10037. Note that our QoI scores are designed to provide relative scores by comparative evaluation instead of an absolute score to aid in ranking.

\begin{figure}
  \includegraphics[width=0.90\linewidth]{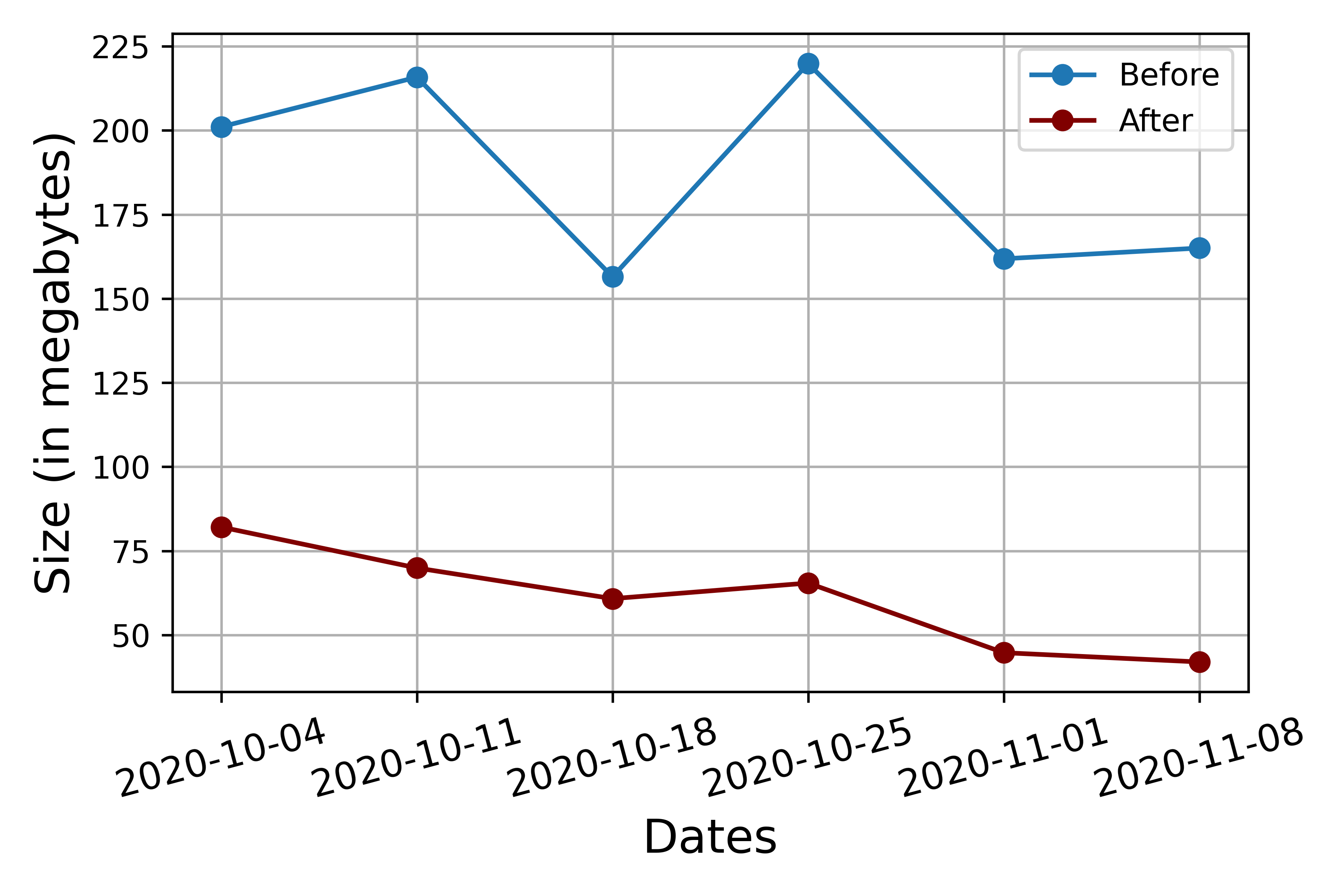}
  \caption{Data set size before and after filtering based on the quality of information in the temporal quality attribute. We use temporal frequency as query criteria to filter images from every Friday between October 5, 2020, to November 15, 2020, which reduces the dataset storage cost by 60\%.}
  \label{fig:storage_before_after}
  \vspace{-10pt}
\end{figure}


\subsection{Practical Use Cases}
We demonstrate here how the proposed framework can address some of the obstacles experienced by urban researchers and planners with large-scale street view images, as revealed in section~\ref{sec:needfinding}. To do this, we present 3 use cases for the proposed QoI framework to incorporate into their current workflow and practices of urban analysis. Towards that goal, we present scenarios where urban experts can gain useful insights about the data cost by querying data using this framework, identifying areas to improve data quality, and focusing the budget to ask for data curated for a specific research goal.

\begin{figure}
\centering
    \includegraphics[width=.49\linewidth]{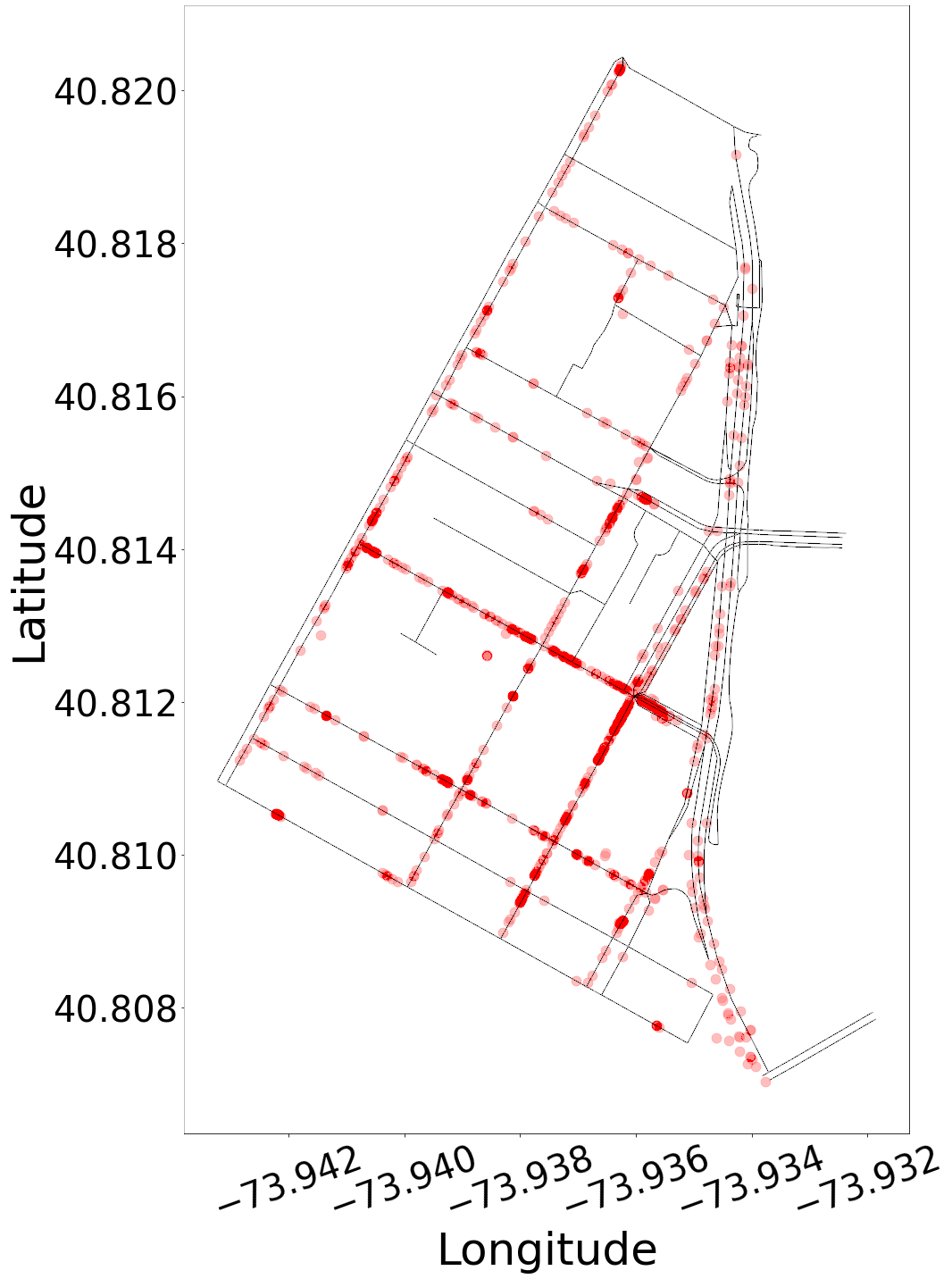}
    \includegraphics[width=.49\linewidth]{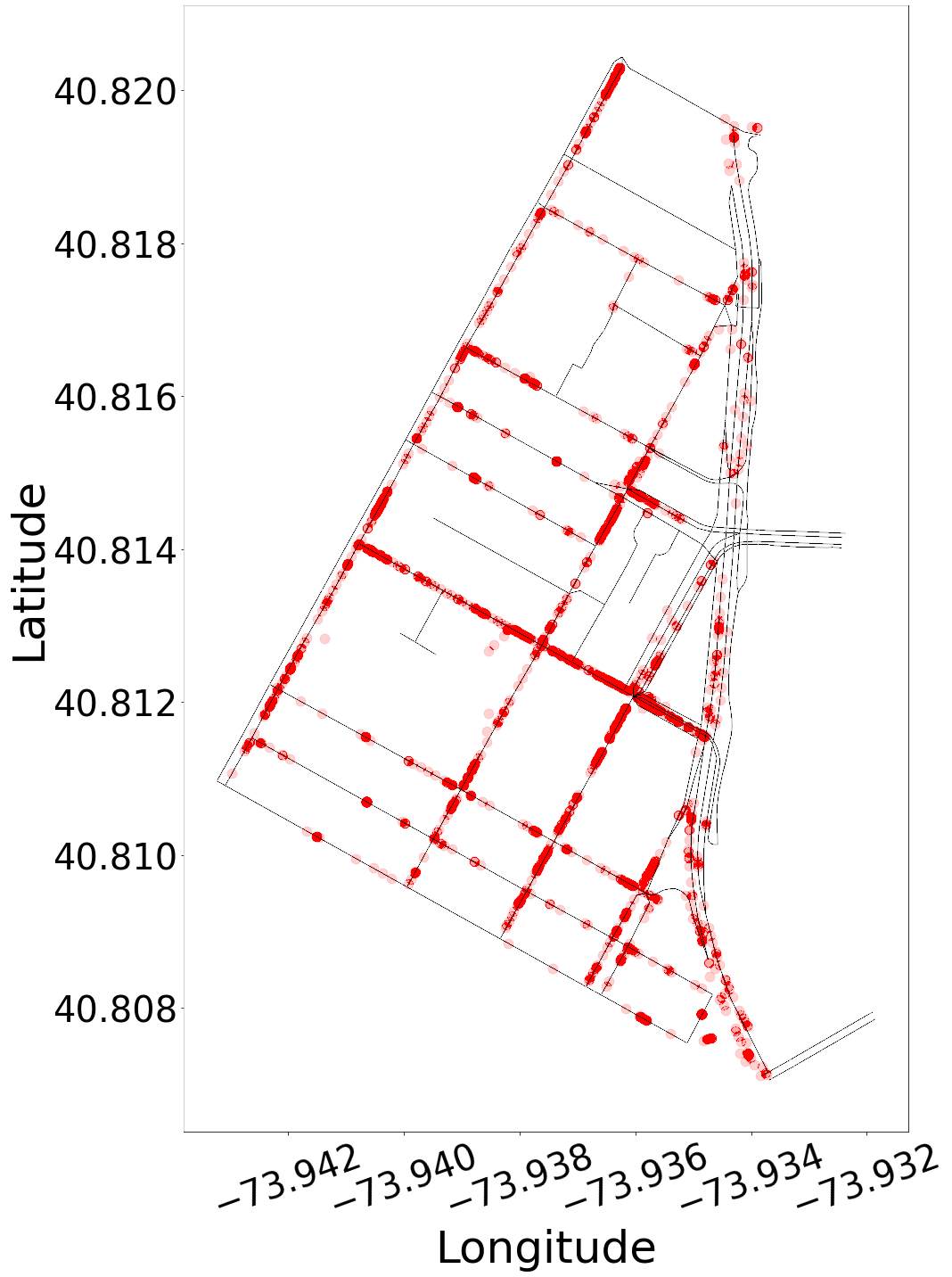}\hfill
    \caption{Example of Spatial data distribution on low (left) and high (right) QoI day for Zip code area 10037.}
\label{fig:high_low_spatial_10037}
\end{figure}

\textbf{1. Assessing Cost.} Our preliminary interview with urban experts revealed how not being able to query and filter data specific to their needs limits their ability to make decisions for procuring the data and calculate the cost associated with this action. This cost involves not only the cost to purchase a data license and store the data but also and labor cost involved in processing and labeling if necessary. The proposed QoI framework can thus be used to focus analysis only on specific attributes and image samples relevant along that dimension.

Consider an urban planning professional who is interested in studying how curbside parking at different times of the day is used in several selected blocks or an entire Zipcode area. For such analysis, the research question needs to explore how roadside parking usage evolves with the time of the day, which involves the temporal dimension of the data. They can evaluate available data for the area of interest based on the temporal quality attribute and decide whether to allocate a budget and purchase the data and storage. As P2 described this decision-making process as `We were evaluating cost as like the value of going through this process and getting proprietary data, and ultimately utilize this data'. To demonstrate how our QoI framework can address this need, we pick the curbside parking problem in a busy city area as an exemplary use case. When we query a subset of the dataset using temporal quality (the more frequently sampled, the better) as query criteria for all six Fridays present in the dataset, dataset size is reduced by 60\%, as we show in Figure~\ref{fig:storage_before_after}. As a result, the organization can make an informed decision about procuring the data segment captured with that criterion. This can enable procuring a smaller volume of data and lower the associated cost by assessing the quality first.

\textbf{2. Improving Quality and Fidelity.} Evaluating the quality of information along specific attributes is the first step for user. For data segment with insufficient quality, identifying areas to sample more in the future to improve the quality is the next step for both urban professionals and data providers. This can potentially improve the data quality over time by identifying missing points that can be referred to as `data deserts'. P1 illustrated this situation as, 'We're always really interested in looking at where there could be a service desert as we're trying to identify where there might be need.
Anything that would help us better pinpoint those deserts I think would be helpful.'

\begin{table}
\begin{tabular}{||c  c c c c c c||}
\multicolumn{1}{c}{} & \multicolumn{6}{c}{\textbf{Zipcode Areas}}\\
\hline
           & 10012 & 10016 & 10018 & 10021 & 10037 & 00083                        \\ \hline
                 \hline
\textbf{Project: Mobility pattern} & 1.278 & 2.519 & 3.014 & 6.638 & 7.715 & 4.038                        \\ \hline

\textbf{Project: Bike usage pattern}  & 1.351 & 1.774 & 6.925 & 8.110 & 6.513 & 2.291                        \\ \hline
\textbf{Project: Air quality estimation} & 1.635 & 2.800 & 6.080 & 9.156 & 8.577 &  4.256 \\ \hline

\end{tabular}
\caption{Quality of Information score for 6 Zipcode areas based on application-specific weighting ($\alpha$, $\beta$, $\gamma$) in eqn. 4.}
\label{tab:composite_ranking_change}
\vspace{-20pt}
\end{table}

For example, a researcher interested in using street view images for urban computing has just found out that the areas their dataset covers have low information quality in certain areas, and using this data will affect their analysis. Figure~\ref{fig:high_low_spatial_10037} shows an example of this scenario, where spatial distribution on a day with high QoI is compared with one with low QoI. Notice the missing points in coverage on a low QoI day (left) as opposed to more dense coverage on a high QoI day (right). Both users and providers need to identify which points need to be sampled to improve their spatial QoI. A similar approach can be used to identify missing points for temporal quality where data needs to be sampled at a higher temporal frequency for the purpose of the analysis. A researcher interested in obtaining data at a high temporal frequency at certain intersections in the street network can locate points that are sampled ten or more times in a day. To improve the temporal frequency of the area, the researcher can then request more data to be sampled along other street segments to improve information quality.


\textbf{3. Use-inspired Data Curation.} The proposed QoI framework is a generic approach that can be customized to specific user needs, as outlined in~\ref{customize_QoI}. By selecting values for the weighting coefficients, users can prioritize different dimensions according to their judgment and obtain a use-specific QoI and ranking that can be expanded across applications.

Consider three researchers interested in using our unified QoI framework for their independent projects. Project 1 involves measuring mobility patterns in specific street intersections at different times of the day (temporal). Project 2 seeks to map bike usage distribution (spatial). Project 3 is interested in estimating greenery and air quality from detected objects in the image scene (content).

The researchers can select weights for individual QoI attributes according to their need and make query decisions based on the result. For illustration purposes, we compute QoI for different combination of values for $\alpha$, $\beta$, and $\gamma$, resulting in a different query criterion for each project aiding in user-inspired data curation. Here, for Project 1, spatial, temporal, and content quality is assigned importance levels of 1 (Not Important), 5 (Very Important), and 3 (Moderately Important). Projects 2 and 3 are assigned importance levels similarly based on the user's organizational and research goal. This allows to address their needs to filter data through a process that incorporates both their disciplinary insights and computational metrics. We present the QoI results for 6 representative zip code areas in Table~\ref{tab:composite_ranking_change}.

\section{Discussion}

\begin{figure}
  \includegraphics[width=0.60\linewidth]{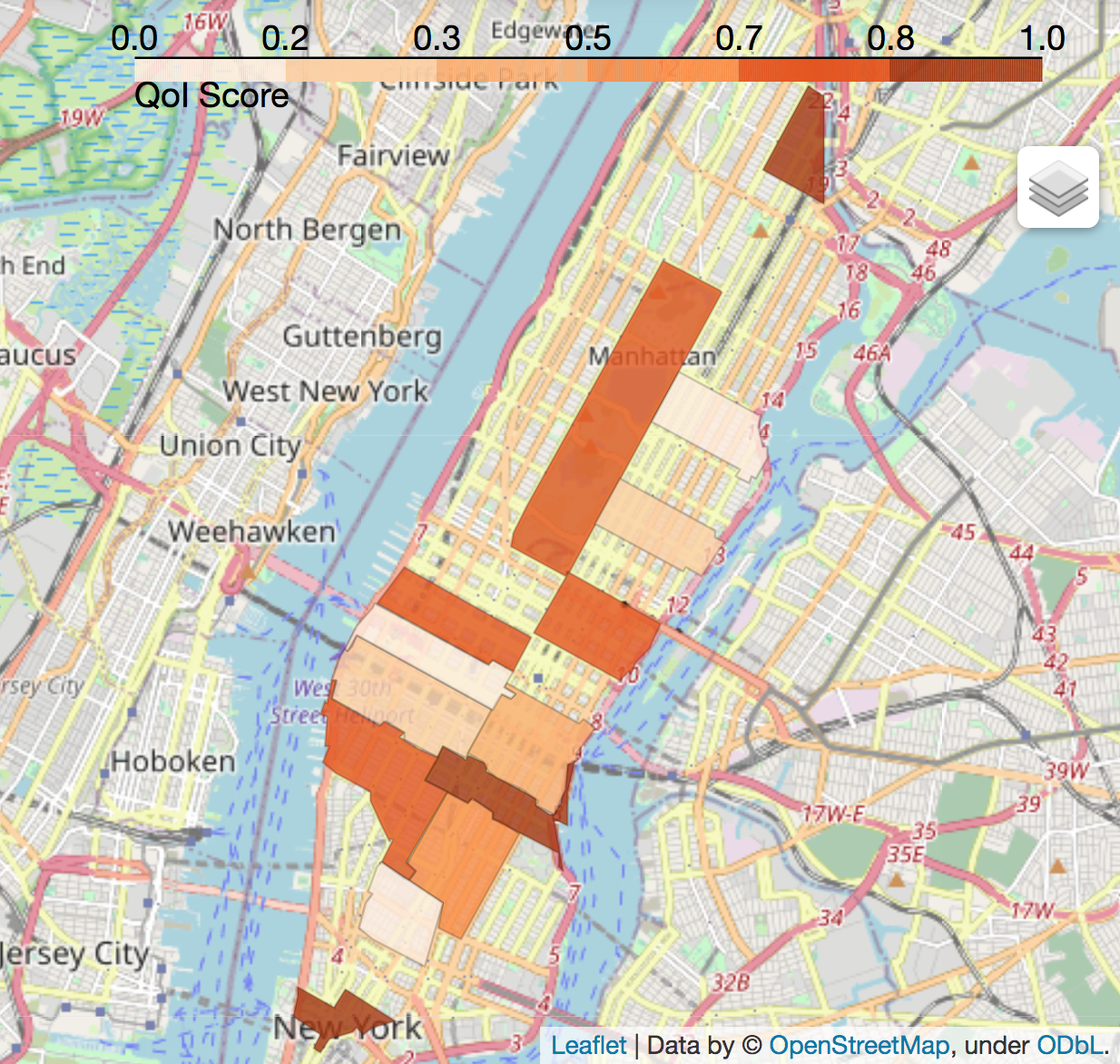}
  \caption{QoI score of 14 zipcode areas. The darker the shade of orange is, the higher information quality is.}
  \label{fig:QoI_14zipcodes}
  \vspace{-10pt}
\end{figure}

Large-scale street view images are complex datasets, yet hold tremendous potential for sensing the urban living and landscapes. Although originated as a concept to map all the streets in the world, the data-driven services offered by the providers in the street view images market offer opportunities to accelerate cataloging aspects of urban quality of life and improve the workflow of urban planners and researchers, as evidenced by prior research. In this work, we focused on understanding the different ways SVI datasets can open new research and insights useful for understanding urban life. Our analysis revealed that while well-known data sources such as Google Street View have been utilized for many use cases, their usability has trade-offs from unreliable update frequency across areas and the cost of procuring data. While there are alternative data providers, the lack of tools and frameworks to support analysis and assessment of dataset utility for certain research goals based on cost and data quality are the key obstacles to better integrate these data in policy design, resource allocation, and services. This work presents a foundation for designing tools to support these needs through a framework for assessing the quality of street view data that is customizable to user preference for a range of use cases. Below we discuss some of the design considerations for such tools and potential design directions for future works.

\subsection{Interaction with Big Data for Exploratory Research}

Large-scale street view images are `big data', both in volume and scale, and interacting with big data is challenging. While there have been some approaches as discussed in section \ref{sec:background_human_centered}, images can be noisy, and require large amounts of storage and processing power for analysis. While artificial intelligence-based techniques for image analysis have improved significantly since the 2010s~\cite{alom2018history}, framing the task for rudimentary automated approaches to extract insights remains difficult when the goal of human inspection is not well-defined, which is often the case in urban analysis. 
Some of the challenges of working with street view images have been observed in the field of satellite imagery observations.
The data is massive, the mode of analysis is often exploratory and may require expert labeling guided by disciplinary knowledge. Both satellite and street view imagery raise persistent privacy concerns that individuals, states, and corporations continue to navigate. 

Some of the interactive tools that have been developed by the remote observation community point towards similar approaches that might be advantageous to the street view imagery datasets as well. Platforms and infrastructure such as Google Earth Engine \cite{gorelick2017google} and Sentinel Hub \cite{milcinski2017sentinel} demonstrate how massive parallel computing can be made easy to use and accessible to end users with less technical experience and can democratize analysis. They also include pricing structures that make non-commercial, academic, and government use of these datasets affordable or free, further incentivizing their use. Accessible and collaborative software for generating custom maps \cite{kastanakis2016mapbox, felt, placemark} is beginning to address some of the needs of community organizations, as described by our participants. Some data providers are beginning to integrate tooling for custom maps, filtering by object detection query, and easier data exploration. While these tools are still limited to predetermined categories with little customizability, working with stakeholders to understand their needs and workflow can minimize the gap between their expectations and existing available tools. 


\subsection{Sustainable Approaches to Augment Human Inspection}
Street View Images capture many nuanced urban scenes, and their interpretation is subject to human observers. Researchers can audit images by manual inspection to codify the physical assets and behavior they observe in the images, but this does not scale. On the other hand, researchers relying on image processing approaches to extract computational measures require pre-existing models trained on similar data for common computer vision tasks (object and people detection, semantic segmentation) or to modify a model on a small set of labeled data labeled for a specific needs. Both approaches have constraints for `out-of-context' scenarios that can lead to information extracted that does not provide a holistic urban scene perception. Recent advances in Large Language Models (LLM) and, trained on vast amounts of labeled training data, have shown good performance for various language tasks including image descriptions. Prior works have used automated image captioning and text detection to annotate street scenes~\cite{chowdhury2021tracking, 9010273},  and LLM's general purpose description capability can be deemed useful for generating annotations and scene descriptions for visual aspects auditing. 
Recent work from computational social science, suggest LLMs may speed up annotation~\cite{ziems2023can}. One strategy could be to augment current human auditing process with `human-in-the-loop' annotation where model generated code is verified and finalized by disciplinary experts.

An important consideration in the use of LLMs for augmenting human annotation will be the environmental cost incurred from their use. Large language models are trained on terabytes of data and are expensive to retrain in terms of energy usage and greenhouse gas emissions. The environmental footprint for large language models is growing quickly over time, which can be a major trade-off in using LLMs for large-scale inference tasks such as annotating millions of street images. As language models like GPT-4's pricing policy changes, financial and environmental costs can influence equitable access to these models for practitioners and institutions already working under smaller budget constraints to achieve urban development goals~\cite{strubell2019energy}.

\subsection{Privacy, Equity, and Policy Considerations}

Most publicly licensed street view imagery (Mapillary and KartaView) have followed Google's precedent to maintain individual privacy, by blurring faces and identifiable information. One of our participants pointed out that SVI datasets can have such high spatial density and image frequency that a person could be tracked via their clothing even if their face is blurred. Removal of all people from images can conflict with other use cases, and further work is necessary to understand the potential privacy harms that high-frequency street view data might pose and avoid `start picking up on patterns and triangulating data'. Furthermore,  there are sensitive sites-- parenthood clinics, dispensaries, bars, playgrounds-- that present a risk to different populations, including minority groups, around which active regulation of data collection may be necessary.

On the other hand, one of our participants suggested the potential use of SVI for security surveillance systems as `if you have good coverage and videos that are updated frequently enough, you can leverage it as a security system for the city'. This indicates that access to large-scale SVI images enables data usage that violates people's civil liberties if potential abuses are insufficiently managed by policy regulations. Policies should be enacted to control the production, sale, and distribution of these datasets and avoid misuse. Existing dataset standardization procedures such as Datasheets for Datasets~\cite{gebru2021datasheets} and Data Statements~\cite{bender-friedman-2018-data} can facilitate better transparency between dataset creators and dataset consumers to encourage accountability.

Ensuring frequent updates of imagery can benefit from broad infrastructural investment and gamified incentive structure for maintaining crowd-contribution in data capture~\cite{quack_2021}. Both KartaView and Mapillary publicly display leaderboards to encourage participation and data collection. However, these incentives are still proving insufficient as significant inequality in data captured from urban and rural areas~\cite{ma2019state}. Prior works in urban sensing systems have explored two paradigms for sensor data collection: opportunistic and participatory data collection, using micro-tasks contribution from crowd~\cite{10.1145/2556288.2556996}, community vehicular networks~\cite{aoki2009vehicle}, and improving coverage of opportunistic data with targeted participatory contributions~\cite{10.1145/3359192}. Similarly, interactive tools for mapping geo-coded data based on different contextual features to probe, visualize, and analyze hypothetical scenarios with minimal coding as ~\cite{10.1145/2556288.2557228, 8807255} can be designed for street view images. Collaborations with urban planners and practitioners to identify proper incentive mechanisms to minimize coverage holes can help guarantee more equitable coverage \cite{yoong_2018}.


\subsection{Limitations and Future Works}

This work is focused on developing a framework for assessing the quality of information of street view image datasets based on findings from an interview study. Our framework provides a basis for designing future interactive tools to support the use of emerging street view datasets in urban planning and research, all aspects of which we were not able to capture within the scope of this study. Though the number of our interview participants was small, based on our thematic analysis, we found the number of study participants to be sufficient to reach convergence. We primarily recruited participants who are familiar with street view image use cases in urban areas like New York City. Our findings and evaluation are based on a single SVI dataset primarily collected in North America. Future studies can investigate how the use cases of SVI vary across different geographic regions and what unique challenges researchers encounter in those contexts. 

The quality assessment framework we present in this work is a general framework and we believe that an end user will find it most effective by adapting it to a specific application. Future works could explore expanding the quality attributes to be adapted to specific user groups and user needs. Due to the lack of ground truth, it is not possible to validate the QoI measures with real-world observations. One approach to validate the QoI measure will be to work with potential stakeholders to assess the utility of the QoI score. Street view images, both crowd-sourced and commercial, often have unequal coverage in geographic areas with low commercial activities and local digital infrastructure, including countries in Africa, and rural areas in Western countries. With increasing use of AI-based image recognition techniques applied to capture the change in landscape and societies, future works need to consider this skewed representation so that policies and research not only benefit higher-income geographic regions but also can scale equitable coverage for developing areas.




\section{Conclusion}
Intersecting trends including excitement over autonomous vehicles, ride-sharing growth, vehicles integrated with cameras, and the increasing availability of cheap and ubiquitous connected devices have resulted in significant growth in the availability of street view imagery. These images vary in licensing, quality, and update frequency, and that influences how end-users experience these datasets changes.
With the overarching goal of supporting geospatial data analysis to improve urban life conditions in a sustainable and equitable way, this work sets out to identify design opportunities centered around the quality, value, and cost of these datasets. We present an overview of the challenges and opportunities these novel datasets might present to urban planners and researchers, to help research and products based on SVI to serve better the needs of people who are working on city services and planning. Building on our findings from interviews with 5 experts from both academia and professional sectors working in this domain, we developed a user-centric information quality metric that allows researchers to assess SVI information quality for spatial, temporal, and content dimensions. Experts interviewed clearly indicated new research opportunities using data with better coverage and frequency that could help establish valuable services to be provided to urban residents. Additionally, new challenges emerge from integrating new datasets, and automated systems into existing workflows, with existing concerns over data equity, access, and privacy. We evaluated our approach using a novel Street View image dataset for practical use cases such as ranking between different data segments, strategic data curation, and query. We propose a single quality metric to allow data filtering based on specific analysis needs and improve data redundancy and sustainability. Moving forward, we believe the findings from this paper will enable the design of customizable, context-dependent, need-based tools to support the works of street view image users.


\bibliographystyle{ACM-Reference-Format}
\bibliography{base, tahiya, new, QoI-ref}

\end{document}